\documentclass[prl,groupedaddress,showpacs,longbibliography,reprint]{revtex4-2}

\usepackage[colorlinks=true,linkcolor=blue,urlcolor=blue,citecolor=blue,pdfusetitle]{hyperref}
\usepackage{color}
\usepackage[utf8]{inputenc}
\usepackage[usenames,dvipsnames]{xcolor}
\usepackage{amsthm}
\usepackage{amsmath}
\usepackage{amssymb}
\usepackage{graphicx}
\usepackage{bm}
\usepackage[all]{xy}
\usepackage{lipsum}
\usepackage{braket}
\usepackage{dsfont}
\usepackage{array}
\usepackage{makecell}
\usepackage{xr}
\usepackage{float}

\newtheorem{theorem}{Theorem}
\newtheorem{proposition}{Proposition}
\newtheorem{corollary}{Corollary}
\newtheorem{lemma}{Lemma}

\begin{document}
\raggedbottom

\title{Non-Gaussian dynamics of quantum fluctuations and mean-field limit in open quantum central spin systems}

\author{Federico Carollo}
\affiliation{Institut f\"{u}r Theoretische Physik,  Universit\"{a}t T\"{u}bingen, Auf der Morgenstelle 14, 72076 T\"{u}bingen, Germany}

\date{\today}

\begin{abstract}
Central spin systems, in which a {\it central} spin is singled out and interacts nonlocally with several {\it bath} spins, are paradigmatic models for nitrogen-vacancy centers and quantum dots. They
show complex emergent dynamics and  stationary phenomena which, despite the  collective nature of their interaction, are still largely not understood. 
Here, we derive exact  results on the emergent behavior of open quantum central spin systems. The latter crucially depends on the scaling of the interaction strength with the bath size. 
For scalings with the inverse square root of the bath size (typical of one-to-many interactions), the system behaves, in the thermodynamic limit, as an open quantum Jaynes-Cummings model, whose bosonic mode encodes the quantum fluctuations of the bath spins. In this case, non-Gaussian correlations are dynamically generated and persist at stationarity. For scalings with the inverse bath size, the emergent dynamics is instead of mean-field type. Our work provides a fundamental understanding of the different dynamical regimes of central spin systems and a simple theory for  efficiently exploring their nonequilibrium behavior. Our findings may become relevant for developing fully quantum descriptions of many-body solid-state devices and their applications. 
\end{abstract}

\maketitle
Collective quantum systems, such as spin ensembles with infinite-range interaction, are ubiquitous in physics and naturally emerge, e.g., in cold-atom experiments   \cite{ritsch2013,norcia2018,dogra2019,muniz2020,mivehvar2021,suarez2023,gabor2023}. 
The broad set of tools available for these systems   \cite{hepp1973,hepp1973b,emary2003,emary2003b,chase2008,baragiola2010,sieberer2016,kirton2017,shammah2018,reitz2022,spohn1980,alicki1983,mori2013,benatti2016,merkli2018,benatti2018,carollo2021} permits for an in-depth characterization of their emergent behavior \cite{dicke1954,hepp1973,hepp1973b,wang1973,hioe1973,carmichael1973,emary2003,sanchez2019,buca2019,reitz2022,tomadin2010,kirton2019,boneberg2022,mattes2023}, which is, in general, exactly described by a {\it mean-field} theory \cite{spohn1980,alicki1983,mori2013,benatti2016,merkli2018,benatti2018,carollo2021,fiorelli2023}. 

\begin{figure}[t!]
    \centering
    \includegraphics[width=\columnwidth]{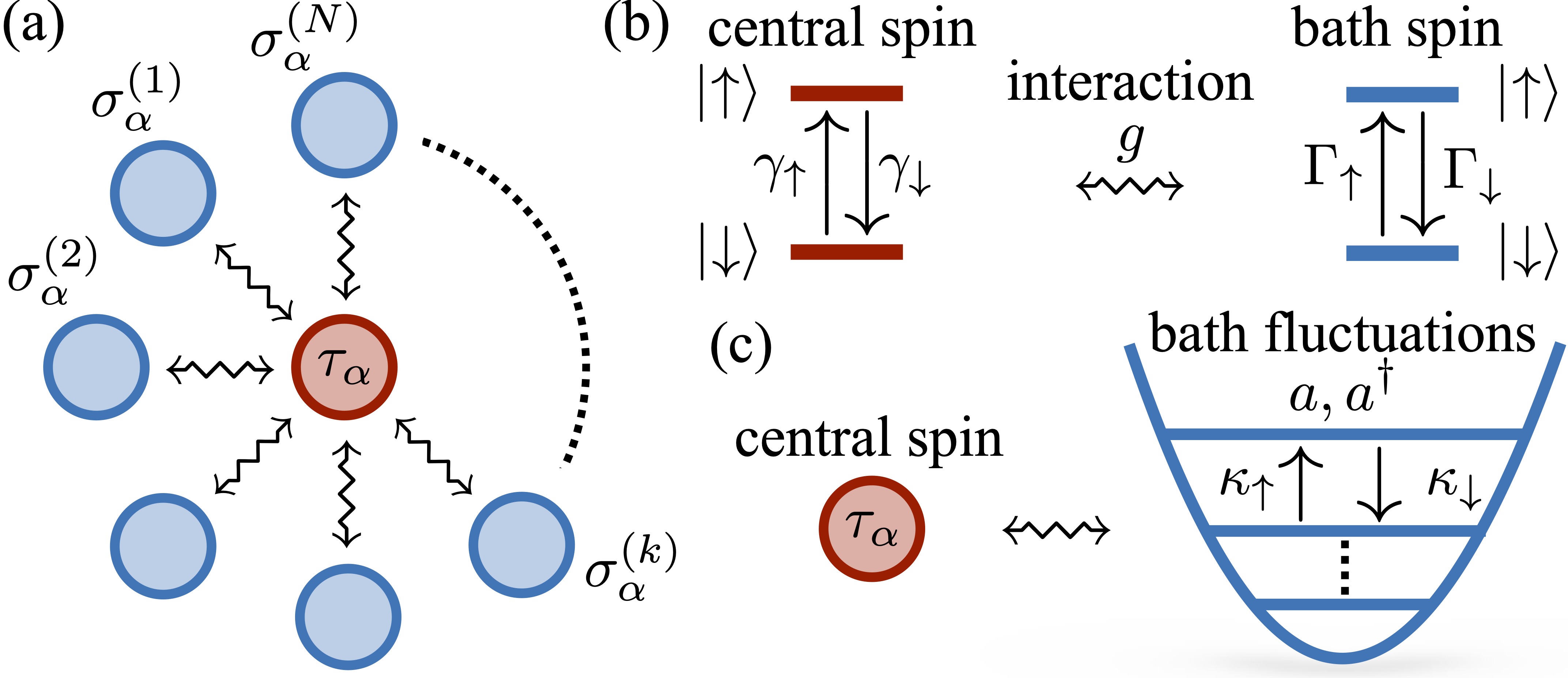}
    \caption{{\bf Sketch of the system.} (a) A  central spin, described by Pauli matrices $\tau_\alpha$, interacts with $N$  bath spins, denoted by the matrices $\sigma_\alpha^{(k)}$. (b) Spins are subject to decay and pump of excitations, with rates $\gamma_{\uparrow,\downarrow}$ ($\Gamma_{\uparrow,\downarrow}$) for the central spin (bath spins). The central spin interacts with the bath spins, with coupling strength $g$, via exchange of excitations. (c) For $g\sim 1/\sqrt{N}$ and in the limit $N\to\infty$, the central spin system behaves as an open quantum Jaynes-Cummings model. The bosonic mode accounts for the quantum fluctuations of the bath spins, which develop a strong non-Gaussian character. }
    \label{Fig1}
\end{figure}

A paradigmatic class of many-body systems featuring collective interaction is that of (open quantum) central spin systems \cite{yuzbashyan2005,bortz2007,bortz2007b,coish2007,maletinsky2009,kessler2010,kessler2012,schwartz2016,gangloff2019,cabot2022,greilich2023}. The latter consist of a {\it central} spin which couples nonlocally  to $N$ {\it bath} spins, with interaction strength $g$ [cf.~Fig.~\ref{Fig1}(a-b)]. These systems provide  quantum models for nitrogen-vacancy centers and quantum dots, and describe their applications as quantum memories or nanoscale quantum sensors \cite{schliemann2003,taylor2003,togan2011,urbaszek2013,aany2017,fernandez2018,villazon2021,rizzato2022,allert2022}. Despite such a broad relevance, an emergent theory for central spin systems in the thermodynamic limit is still missing \cite{yang2008,urbaszek2013,chekhovich2013,lindoy2018,rohrig2018}, especially  within the framework of open quantum systems  \cite{kessler2010,kessler2012,fowler2023}. In this regard, a key complication arises from the fact that, even though they feature a collective interaction [cf.~Fig.~\ref{Fig1}(a)], central spin systems are not always captured by a mean-field theory \cite{fowler2023}. This observation poses the  challenge of understanding why mean-field theory can fail to describe these systems in certain parameter regimes and whether there still exists an effective theory for these cases. Answering these questions can pave the way to  a fully quantum description of many-body solid-state devices \cite{yang2008,urbaszek2013,chekhovich2013,lindoy2018,rohrig2018}, thus enabling the analysis and the exploration of protocols for controlling quantum bath-spin many-body dynamics or for engineering correlated quantum states \cite{gangloff2019,rudner2011,delange2012,gao2015,bauch2018,chen2018,unden2018,dong2019,liu2021,degen2021,gillard2022}.

In this paper, we make progress in this direction by analytically deriving the emergent dynamical theory for  open quantum central spin systems [cf.~Fig.~\ref{Fig1}(a-b)]. For $g\sim 1/\sqrt{N}$, we show that the central spin system behaves, in the thermodynamic limit, as a one-spin one-boson system, related to the  Jaynes-Cummings model \cite{jaynes1963}, which encodes the coupling of the central spin with the quantum fluctuations \cite{goderis1989,goderis1990,benatti2017,benatti2018} of the bath [see illustration in Fig.~\ref{Fig1}(c)]. In this scenario, the system does not obey a mean-field theory, but  rather a quantum {\it fluctuating-field} one, and develops strong long-lived non-Gaussian correlations which persist in the thermodynamic limit. Central spin systems are thus a promising resource for engineering complex quantum fluctuations and non-Gaussian correlations in many-body systems---which is a matter of current interest \cite{stitely2023}.
We further consider an interaction strength scaling as $g\sim 1/N$. In this case, the central spin couples to the average  behavior of the bath spins and the system is described by a mean-field theory. 

Our work delivers new insights into the dynamics of open quantum central spin systems and resolves a discrepancy between recent numerical results and mean-field prediction for these systems \cite{fowler2023}.  
It further provides a clear-cut example of a quantum fluctuating-field theory in open quantum systems, whose fundamental properties may be relevant for developing an emergent theory for  solid-state devices, able to account for many-body spin baths in the quantum regime    \cite{yang2008,urbaszek2013,chekhovich2013,lindoy2018,rohrig2018}.  \\

\noindent \textbf{Central spin system.---} We focus on the system depicted in Fig.~\ref{Fig1}(a), consisting of $N+1$ spin-$1/2$ particles, with basis states $\ket{\uparrow}$ and $\ket{\downarrow}$. The central spin is described by the Pauli matrices $\tau_\alpha$ while $\sigma_\alpha^{(k)}$, $k=1,2,\dots N$ are those of the bath spins. The system Hamiltonian is (see below for extensions)
\begin{equation}
H=H_\tau+H_{\rm int}\, ,\quad \mbox{with} \quad \, H_{\rm int}=g\left(\tau_+S_-+\tau_-S_+\right)\, .
\label{Hamiltonian}
\end{equation}
Here, $H_\tau=\sum_\alpha w_\alpha \tau_\alpha$ is the Hamiltonian of the central spin only, $S_\pm=\sum_{k=1}^N\sigma_\pm^{(k)}$ and $\tau_\pm,\sigma_\pm$ are ladder operators, e.g., $\sigma_-=\ket{\downarrow}\!\bra{\uparrow}$ and $\sigma_+=\sigma_-^\dagger$. The interaction Hamiltonian $H_{\rm int}$ describes a collective excitation exchange, with coupling strength $g$, between the bath spins and the central one [cf.~Fig.~\ref{Fig1}(a-b)].
The system is also subject to irreversible processes, shown in Fig.~\ref{Fig1}(b), so that the dynamics of any system operator $X$ is implemented by the  equation $\dot{X}_t=\mathcal{L}[X_t]:=i[H,X_t]+\mathcal{D}_\tau[X_t]+\mathcal{D}_{\rm bath}[X_t]$ \cite{lindblad1976,gorini1976,breuer2002theory}, where 
\begin{equation}
\begin{split}
\mathcal{D}_\tau[X]&=\gamma_\downarrow\mathcal{W}_{\tau_-}[X]+\gamma_\uparrow\mathcal{W}_{\tau_+}[X] \, ,\\
\mathcal{D}_{\rm bath}[X]&=\sum_{k=1}^N\left(\Gamma_\downarrow\mathcal{W}_{\sigma_-^{(k)}}[X]+\Gamma_\uparrow \mathcal{W}_{\sigma_+^{(k)}}[X]\right)\, .
\end{split}
\label{dissipator}
\end{equation} 
The rates $\gamma_{\uparrow,\downarrow}$  ($\Gamma_{\uparrow,\downarrow}$) are associated with irreversible pump and decay of excitations for the central spin (bath spins) and $\mathcal{W}_\nu[X]=\nu^\dagger X \nu-(\nu^\dagger \nu X+X\nu^\dagger \nu)/2$.

A particular instance of the system above was investigated  in Ref.~\cite{fowler2023}. It was numerically shown that a mean-field approach, obtained by neglecting correlations among spins, does not capture the behavior of the system in the thermodynamic limit, for $g\sim1/\sqrt{N}$. This came as quite a surprise since it is, at least at first sight, in stark contrast with what happens to structurally similar spin-boson models \cite{tavis1967,tavis1969,kirton2017,kirton2019,carollo2021,fowler2023}. In what follows, we rigorously explain the dynamical behavior of  central spin systems through exact analytical results.

Since we will work in the limit $N\to\infty$, it is useful to make a few considerations on the generator $\mathcal{L}$. Its dissipative terms describe irreversible processes occurring independently for each spin and are thus well-defined for any $N$. The Hamiltonian $H_{\rm int}$ in Eq.~\eqref{Hamiltonian} shows instead a peculiar behavior. From the viewpoint of the bath spins, it features the expected extensive character, with norm proportional to $gN$. However, this extensivity is problematic for the central spin. To see this, let us compute 
\begin{equation}
\Omega\left([H_{\rm int },\tau_{z}]\right)=2gN \left(\tau_- \langle \sigma_+\rangle -\tau_+ \langle \sigma_-\rangle\right)\, ,
\label{cn1}
\end{equation}
where $\Omega$ is the partial expectation over the bath spins, such that $\Omega(\tau_- S_+)=\tau_-\langle S_+\rangle$, which we assume to be uncorrelated, $\Omega(\sigma_\alpha^{(k)}\sigma_\beta^{(h)})=\Omega(\sigma_\alpha^{(k)})\Omega(\sigma_\beta^{(h)})$, $\forall k\neq h$, and permutation invariant, $\Omega(\sigma_\alpha^{(k)})=\langle \sigma_\alpha\rangle$, $\forall k$. 
In the thermodynamic limit, Eq.~\eqref{cn1}, which provides a term appearing in the time-derivative of $\tau_z$ at time $t=0$, diverges unless $\langle \sigma_\pm\rangle=0$. Even using this assumption, the term 
\begin{equation}
\begin{split}
\!\!\Omega\left([H_{\rm int}, [H_{\rm int}, \tau_z]]\right)
=4 g^2 N \left(\tau_z\langle \sigma_+\sigma_-\rangle -\tau_{+}\tau_- \langle \sigma_z\rangle \right)\, ,
\label{cn2}
\end{split}
\end{equation} 
shows that the Heisenberg equations for the central spin can diverge with $N$. To make the above dynamics well-behaved, one has to appropriately rescale $g$. We first consider $g\sim 1/\sqrt{N}$ and show that this choice gives rise to an effective one-spin one-boson dynamics. Later, we turn to the case $g\sim 1/N$ which, as we demonstrate, is instead exactly described by a mean-field theory. \\

{\noindent }{\bf Local state of the bath spins.---} Rescaling the coupling constant also affects the dynamics of the bath spins. Considering a generic {\it local} bath operator $A$ (i.e., an operator solely acting on a finite number of bath spins \cite{SM}\vphantom{\cite{bratteli1982,thirring2013,strocchi2021,verbeure2010,benatti2015}}), we indeed have that $\left\|[H_{\rm int},A]\right\|\sim g$ vanishes whenever $g$ decays with $N$. 
This implies that the Hamiltonian $H_{\rm int}$ is irrelevant for the dynamics of local bath operators, which thus solely evolve according to $\mathcal{D}_{\rm bath}$ in the  thermodynamic limit. This fact is summarized in the following Lemma, whose proof is given in Ref.~\cite{SM}.
\begin{lemma}
\label{local-bathspin}
For an interaction strength $g=g_0/N^{z}$, with $z>0$ and $g_0$ an $N$-independent constant, we have
$$
\lim_{N\to\infty}\left\|e^{t\mathcal{L}}[A]-e^{t\mathcal{D}_{\rm bath}}[A]\right\|=0\, , 
$$ 
for any local bath-spin operator $A$. 
\end{lemma}

The time evolution of local operators of the bath spins, e.g., the operators $\sigma_\alpha^{(k)}$, and of the so-called average operators  $m_\alpha^N=\sum_{k=1}^N \sigma_\alpha^{(k)}/N$ as well (see Ref.~\cite{SM}), is thus not affected by the presence of the central spin. Furthermore, the dynamics generated by $\mathcal{D}_{\rm bath}$ drives the bath spins towards the permutation-invariant uncorrelated state $\Omega_{\rm SS}$, defined by the expectation values $\Omega_{\rm SS}(\sigma_z)=\Gamma_-/\Gamma_+$, with $\Gamma_\pm=\Gamma_\uparrow\pm\Gamma_\downarrow$, and $\Omega_{\rm SS}(\sigma_\pm)=0$. Note that the latter relation, combined with the rescaling $g\sim 1/\sqrt{N}$, gives a well-defined thermodynamic limit for Eqs.~\eqref{cn1}-\eqref{cn2}.  It is thus reasonable to assume $\Omega_{\rm SS}$ to be  the ``reference" (initial) state for the bath spins. 
As we shall see below, the bath spins nevertheless experience some dynamics. Their quantum fluctuations, described by nonlocal unbounded operators, are indeed affected by the coupling with the central spin and thus can evolve in time \cite{narnhofer2002,benatti2016,benatti2018,merkli2018,carollo2022}.  Without loss of generality, we focus on the case $\Gamma_\uparrow<\Gamma_\downarrow$ and define $\varepsilon:=-\Omega_{\rm SS}(\sigma_z)>0$. \\

{\noindent }{\bf Bath-spin fluctuations.---} For $g=g_0/\sqrt{N}$, the central spin couples to bath operators of the form $S_\pm/\sqrt{N}$, as clear from Eq.~\eqref{Hamiltonian}. These nonlocal unbounded operators are known as quantum fluctuation operators and behave, in the thermodynamic limit, as bosonic operators \cite{goderis1989,goderis1990,matsui2003,verbeure2010,benatti2017}. This can be understood by considering their commutator $[S_-,S_+]/N=-m_z^N$, which is proportional to an average operator. For product states like $\Omega_{\rm SS}$, average operators converge to their expectation value \cite{landford1969,bratteli1982,thirring2013,strocchi2021}, essentially due to a  law of large numbers. As such, we have $[S_-,S_+]/N\to \varepsilon$, which suggests the definition of the rescaled quantum fluctuations
\begin{equation}
a_N=\frac{S_-}{\sqrt{\varepsilon N}}\, ,  \quad a_N^\dagger =\frac{S_+}{\sqrt{\varepsilon N}}\, . 
\label{FO}
\end{equation}
The latter are such that $[a_N,a_N^\dagger]\to 1$, and thus behave as annihilation and creation operators. The quantum state of the limiting fluctuation operators $a=\lim_{N\to\infty}a_N$ and $a^\dagger=\lim_{N\to\infty}a_N^\dagger$ (where convergence is meant in a quantum central limit sense \cite{goderis1989,goderis1990,benatti2017,SM}) emerges from the state $\Omega_{\rm SS}$. It is a bosonic thermal state $\rho_\beta$ identified by the occupation ${\rm Tr}(\rho_\beta a^\dagger a )=\lim_{N\to\infty}\Omega_{\rm SS}(a^\dagger_Na_N)=\Gamma_\uparrow/(\varepsilon \Gamma_+)$, as proved in the following Proposition. 

\begin{proposition}
\label{QCLT}
The state $\Omega_{\rm SS}$ and the operators $a_N,a_N^\dagger$ give rise, in the limit $N\to\infty$, to a bosonic algebra, with operators $a,a^\dagger$ and state $\tilde{\Omega}_\beta(\cdot)={\rm Tr}\left(\rho_\beta \cdot\right)$, where 
$$
\rho_\beta=\frac{e^{-\beta\omega a^\dagger a}}{1-e^{-\beta\omega}}\, ,  \quad \mbox{ and } \quad \beta\omega=-\log \frac{\Gamma_\uparrow}{\Gamma_+ \varepsilon +\Gamma_\uparrow} \, . 
$$
\end{proposition}
{\it Proof:} Following, e.g., Refs.~\cite{goderis1989,benatti2015}, in order to show that the operators $a_N,a_N^\dagger$ behave, in the limit $N\to\infty$, as bosonic operators equipped with the state $\rho_\beta$, we need to show that (in the spirit of a central limit theorem)
$$
\lim_{N\to\infty}\Omega_{\rm SS}\left(e^{s a_N-s^* a_N^\dagger }\right)=e^{-{|s|^2}/{(2\varepsilon)}}={\rm Tr}\left(\rho_\beta e^{sa-s^*a^\dagger}\right)\, ,
$$
and analogous relations for products of the above exponentials. These limits define an equivalence relation between bath-spin fluctuations and a Gaussian bosonic system. 
The explicit calculation is reported in Ref.~\cite{SM}.  \qed

Mapping bath-spin fluctuations onto bosonic operators shows that the central spin system becomes, in the thermodynamic limit, a one-spin one-boson model [cf.~Fig.~\ref{Fig1}(c)]. The task is now to derive its dynamics.  \\

\noindent {\bf Emergent non-Gaussian dynamics.---}  The terms in the generator concerning the central spin only, i.e., $H_\tau$ and $\mathcal{D}_{\tau}$, are not affected by the limit $N\to\infty$. However, to identify the emergent dynamics we also have to control the action of the generator $\mathcal{D}_{\rm bath}$, and of the interaction Hamiltonian $H_{\rm int}$, on the relevant operators. The aim is then to interpret this action as that of a dynamical generator for the one-spin one-boson model formed by the central spin and the bath-spin quantum fluctuations. 

\begin{figure*}[t!]
    \centering
    \includegraphics[width=\textwidth]{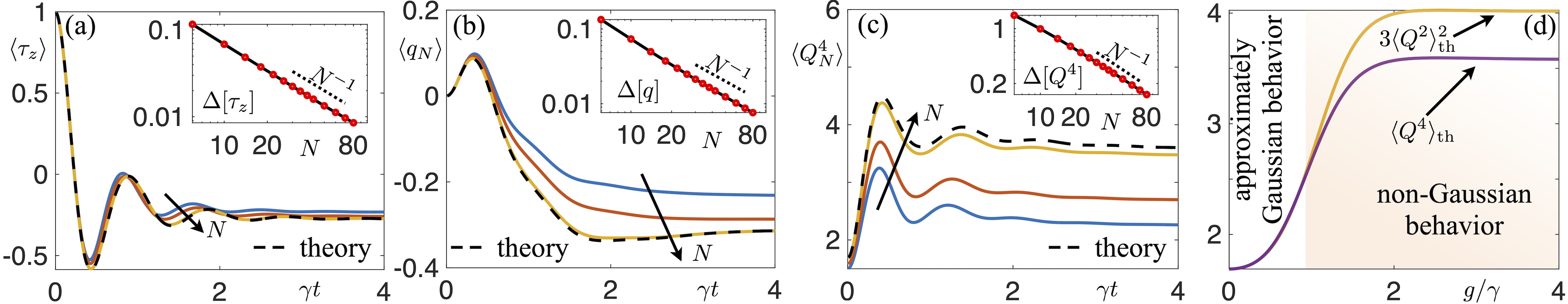}
    \caption{{\bf Emergent dynamics and non-Gaussian fluctuations.} System with $\gamma_\uparrow=0.8\gamma$, $\gamma_\downarrow=0.1\gamma$, $\Gamma_\uparrow=0.2\gamma$, $\Gamma_\downarrow=\gamma$, $w_x=2w_z=\gamma$, and $\gamma$ being a reference rate. Initially, the central spin is in state $\ket{\uparrow}$ and the bath spins are described by $\Omega_{\rm SS}$.  (a) Dynamics of the magnetization $\langle \tau_z\rangle$. The curves shown are for $N=6,10,80$. The dashed line is the model in Eq.~\eqref{L_tilde}. Here, $g/\gamma=4$. The inset shows $\Delta[\tau_z]:=\max_{\gamma t\in[0,4]}|\langle \tau_z\rangle -\langle \tau_z\rangle_{\rm th}|$, where $\langle \cdot \rangle_{\rm th}$ denotes the expectation for the model in Eq.~\eqref{L_tilde}.   (b) Same as (a) for the ``quadrature" $q_N=(a_N+a_N^\dagger)/\sqrt{2}$. The inset shows $\Delta[q]:=\max_{\gamma t\in[0,4]}|\langle q_N\rangle -\langle q\rangle_{\rm th}|$, with $q=(a+a^\dagger)/\sqrt{2}$.  (c) Fourth moment of the centered quadrature $Q_N=q_N-\langle q_N\rangle$,  compared with the prediction (dashed line). The inset displays $\Delta[Q^4]:=\max_{\gamma t\in[0,4]}|\langle Q_N^4\rangle -\langle Q^4\rangle_{\rm th}|$, with $Q=q-\langle q\rangle_{\rm th}$.  (d) Stationary value of $\langle Q^4\rangle_{\rm th}$ compared with the Gaussian estimate $3\langle Q^2\rangle_{\rm th}^2$. The shaded region highlights the regime in which $\langle Q^4\rangle_{\rm th}$ is signalling a strongly non-Gaussian quantum state.  }
    \label{Fig2}
\end{figure*}

For the interaction Hamiltonian, we observe that [recalling Eq.~\eqref{FO} and using that $g=g_0/\sqrt{N}$]
\begin{equation}
i[H_{\rm int }, a_N]= i g_0 \sqrt{\varepsilon} \tau_- [a_N^\dagger,a_N]\, \xrightarrow{N\to\infty} \, -i g_0 \sqrt{\varepsilon} \tau_-\, , 
\label{H_int_TL}
\end{equation}
which follows from the bosonic character of quantum fluctuations. Since central spin operators are not affected by the limit, we conclude that the emergent interaction is described by the Jaynes-Cummings Hamiltonian \cite{jaynes1963}
\begin{equation}
\label{emer_H}
\tilde{H}_{\rm int}=g_0\sqrt{\varepsilon}(\tau_-a^\dagger +\tau_+ a)\, .
\end{equation}
The dissipator $\mathcal{D}_{\rm bath}$ is not of collective type. Still, we can understand its limiting behavior by analyzing its action on the operators $a_N,a_N^\dagger$. We observe that $\mathcal{D}_{\rm bath}[a_N]=-\Gamma_+ a_N/2$, and that
\begin{equation*}
\mathcal{D}_{\rm bath}[a_N^\dagger a_N]=-\Gamma_+ a_N^\dagger a_N+\frac{\Gamma_\uparrow}{\varepsilon}\, \xrightarrow{N\to\infty}  \, -\Gamma_+a^\dagger a+\frac{\Gamma_\uparrow}{\varepsilon}\, .
\label{num_op}
\end{equation*} 
These relations suggest that the dissipative processes are implemented on  quantum fluctuations by the map 
\begin{equation}
\label{emer_D}
\tilde{\mathcal{D}}[X]=\kappa_\downarrow\mathcal{W}_a[X]+\kappa_\uparrow\mathcal{W}_{a^\dagger}[X]\, .
\end{equation}
The above  is a quadratic map and encodes loss and pump of bosonic excitations, with rates $\kappa_\downarrow= \Gamma_++\Gamma_\uparrow/\varepsilon$ and  $\kappa_\uparrow= \Gamma_\uparrow/\varepsilon$ [cf.~Fig.~\ref{Fig1}(c)]. Our considerations, gathered in the following Theorem, allow us to establish that the central spin system becomes an emergent spin-boson system associated with the dynamical generator 
\begin{equation}
\tilde{\mathcal{L}}[X]=i[H_\tau+\tilde{H}_{\rm int}, X]+\mathcal{D}_\tau[X] +\tilde{\mathcal{D}}[X]\, .
\label{L_tilde}
\end{equation}
In essence, this generator describes Jaynes-Cummings physics \cite{jaynes1963} in the presence of dissipation and of a possible Hamiltonian ``driving", $H_\tau$, on the central spin. 
\begin{theorem}
\label{Theo1}
For $g=g_0/\sqrt{N}$, the action of $\mathcal{L}$ on monomials of bath-spin fluctuations and central spin operators gives rise, under any expectation taken with $\Omega_{\rm SS}$, to the map $\tilde{\mathcal{L}}$ on the emergent one-spin one-boson system. 
\end{theorem}
{\it Proof:} The idea is to make the above argument valid for generic monomials of the form $P_N=\tau_\alpha a_N^{\dagger\, \ell} a_N^{k} m_z^{N\, h}$. Due to Proposition \ref{QCLT}, $P_N$ converges to $P=\tau_\alpha a^{\dagger\, \ell} a^k (-\varepsilon)^{h}$ in a ``weak" sense, i.e., whenever considering expectation values constructed with the state $\Omega_{\rm SS}$ and other monomials (see  Ref.~\cite{SM}). This convergence provides the starting point to investigate the action of $\mathcal{L}$. It can indeed be shown that $\mathcal{L}[P_N]$ produces a linear combination of monomials of the same type of $P_N$, plus corrections of order $O(1/N)$ under the considered expectation. Proposition \ref{QCLT} thus guarantees that $\lim_{N\to\infty}\mathcal{L}[P_N]$ converges to a linear combination of monomials of the same form of $P$. By direct calculation, we can show that such linear combination is equal to that produced by $\tilde{\mathcal{L}}[P]$ \cite{SM}. \qed

The theorem directly implies that the dynamics of the emergent one-spin one-boson model, describing the central spin system in the thermodynamic limit, is governed by the generator in Eq.~\eqref{L_tilde},  under physical  regularity conditions on the evolution (see details in Ref.~\cite{SM}). 

To concretely benchmark our derivation, we perform numerical simulations of the model in Eq.~\eqref{L_tilde}. We consider the dynamics of the central spin system, as described by Eqs.~\eqref{Hamiltonian}-\eqref{dissipator}, and analyze convergence of the numerical data for finite systems \cite{chase2008,baragiola2010,kirton2017,shammah2018}    to our prediction, upon increasing the size of the bath. The convergence behavior is shown in Fig.~\ref{Fig2}(a-b-c) for different observables. In the insets of Fig.~\ref{Fig2}(a-b-c), we provide the maximal absolute difference between finite-$N$ results and our prediction for the thermodynamic limit. This error measure decays as $\sim 1/N$, as anticipated in the proof of Theorem \ref{Theo1}, thus confirming the validity of our theory. As shown in Fig.~\ref{Fig2}(d), quite remarkably, the central spin system features, in the thermodynamic limit, non-Gaussian  correlations among the bath spins which persist at stationarity. \\

\noindent {\bf Mean-field regime.---} We now turn to the case $g=g_0/N$. Here, the norms of $H_\tau$ and $H_{\rm int}$ are of the same order and the central spin couples to the (bounded) bath-spin average operators $m_\alpha^N$ [cf.~Eq.~\eqref{Hamiltonian}]. Thus, it is not necessary to require $\langle \sigma_{\pm}\rangle=0$ for a well-defined thermodynamic limit [cf.~Eq.~\eqref{cn1}] and we can therefore consider more involved bath-spin dynamics. For concreteness, we still focus on the dissipator $\mathcal{D}_{\rm bath}$ and introduce a noninteracting Hamiltonian $H_{\rm bath}=\sum_{\alpha}h_\alpha\sum_{k=1}^N \sigma_\alpha^{(k)}$. Other collective dynamics \cite{benatti2018,fiorelli2023} would give analogous results. 

The bath-spin dynamics is not affected by the central spin [cf.~Lemma \ref{local-bathspin}] and the evolved average operators $e^{t\mathcal{L}}[m_\alpha^N]$ converge (weakly) to the time-dependent multiples of the identity $m_\alpha(t)$, obeying a mean-field theory. The central spin instead feels the presence of the bath spins via a coupling to their average operators. The latter thus provide time-dependent (mean) fields ``modulating" the central spin Hamiltonian (see also Ref.~\cite{davies1973}). This is the content of the next Theorem proved in Ref.~\cite{SM}. 

\begin{theorem}
\label{Theo2}
Consider $g=g_0/N$, an initial bath-spin  permutation-invariant uncorrelated state $\Omega$ and the generator $\mathcal{L}$, with  $H\to H+H_{\rm bath}$.  Under any possible expectation (that is, in the weak operator topology \cite{bratteli1982,thirring2013,strocchi2021}), the dynamics of central spin operators is generated, for  $N\to\infty$, by $\mathcal{D}_\tau$ and the time-dependent Hamiltonian 
$$
H_{\tau}^{\rm mf}=H_\tau + g_0\left[m_-(t) \tau_++m_+(t)\tau_-\right]\, .
$$
Here, $m_\pm(t)=[m_x(t)\pm i m_y(t)]/2$ and $m_\alpha(t)$ are the (scalar) limits of the evolved bath-spin average operators. 
\end{theorem}
The exactness of a mean-field theory  in many-body systems is thus not merely related to the structure of the interaction. It relies on substituting certain time-evolved operators, in a finite set, with their expectation value. For this, it is sufficient that: i) the substitution is valid for the initial state, in the thermodynamic limit \cite{landford1969}; ii) the action of the generator on these operators gives a  ``regular" function of them, plus at most terms vanishing with $N$ \cite{benatti2018,carollo2021,fiorelli2023}. If this happens, the involved operators converge to scalars at all times \cite{carollo2021,fiorelli2023,SM}.
Despite the collective interaction, for $g\sim 1/\sqrt{N}$, the central spin couples to the quantum fluctuations of the bath spins, which do not even converge to scalar quantities in the initial state. In this case, mean-field theory cannot be exact. \\

\noindent {\bf Inhomogeneous coupling and extensions.---} Our approach remains valid for inhomogeneous couplings \cite{urbaszek2013,rovnyak2008,villazon2021}. 
To show this, let us consider the interaction 
$
H_{\rm int}=g\tau_+\sum_{k=1}^Nc_k \sigma_-^{(k)}+{\rm h.c.}
$
and set, without loss of generality, $\sum_{k=1}^N|c_k|^2/N\to 1$, in the thermodynamic limit. By defining the quantum fluctuation $a_N=(1/\sqrt{\varepsilon N})\sum_{k=1}^N c_k \sigma_-^{(k)}$, we have that $[a_N,a_N^\dagger]\to 1$, we can prove Proposition \ref{QCLT} and the results in Eqs.~(\ref{H_int_TL}-\ref{L_tilde}), leading to the emergent one-spin one-boson theory for $g\sim 1/\sqrt{N}$. By defining the average bath-spin operators  $m_\alpha^N=(1/N)\sum_{k=1}^Nc_k \sigma_\alpha^{(k)}$ and following Theorem \ref{Theo2}, we can show the validity of mean-field theory for $g\sim 1/N$.

Our derivation holds for generic couplings of the form $\tau_{\alpha}\sum_{k=1}^N \sigma_{x/y}^{(k)}$ and interactions among bath spins (see Ref.~\cite{SM}). It also holds for arbitrary spin particles \cite{benatti2018} and for non-Markovian dynamics with time-dependent generators. In the latter case, time-dependent bath-spin  rates would lead to a time-dependent state $\Omega_{\rm SS}$, and thus to a time-dependent  $\varepsilon$.    
Note that our results could not be obtained via Holstein-Primakoff approaches  \cite{holstein1940}, due to the presence of local dissipation and/or inhomogeneous coupling \cite{kirton2019}. Even when Holstein-Primakoff transformations can be applied (e.g., unitary dynamics \cite{yuan2007,dehghani2020}), our derivation inherently accounts for the state-dependent emergent commutation relation between the operators $S_\pm/\sqrt{N}$ \cite{goderis1989,goderis1990,verbeure2010,benatti2017,benatti2018}, encoded in $\varepsilon$, which may be overlooked by other approaches.     
\\

\noindent {\bf Discussion.---} 
Central spin systems can be realized with nitrogen-vacancy centers or quantum dots \cite{rovnyak2008,kessler2012,kessler2010,fernandez2018,villazon2021}. In these cases, the coupling between the central spin and each bath spin depends on their distance and on the angle between the two spins and the applied magnetic field \cite{villazon2021,rovnyak2008}. This allows one to control the ``microscopic" couplings, and even to realize the $1/\sqrt{N}$ or the $1/N$ scalings, by engineering  suitable structures [e.g., (quasi) one-dimensional ones] and choosing appropriate field directions. In the case of  fixed couplings, desired regimes may instead be achieved by scaling-up with $N$ other parameters, such as the driving fields \cite{kessler2012,fowler2023}.  

Coupling strengths  $g\propto 1/N$, related to the mean-field limit, are accurate in regimes with delocalized central-spin wave function \cite{urbaszek2013}. Still, due to the finiteness of these systems,  bath-spin fluctuations become relevant, on long time-scales, also in these cases    \cite{coish2007,urbaszek2013}.  Our approach allows us to treat them in the quantum regime. Furthermore, central spin systems can be realized with Rydberg atoms \cite{ashida2019,anikeeva2021,dobrzyniecki2023}, guaranteeing highly  controllable couplings \cite{anikeeva2021,dobrzyniecki2023}. Our findings thus also provide a simple way to benchmark these quantum-simulation platforms. 

We now comment on related  results. Refs.~\cite{kessler2012,fowler2023} consider Hamiltonian $H_\tau$ and/or dissipation $\mathcal{D}_\tau$ with the same extensivity of $H_{\rm int}$. This case is similar to that of Theorem \ref{Theo2} and indeed shows mean-field behavior. 
Another related result is Lemma 1.5 of Ref.~\cite{merkli2018}, which focusses on closed systems  with  Hamiltonian given by $H_\tau$ and, e.g., $H_{\rm int }=(g_0/\sqrt{N})\tau_xS_x$. There, the emergent bosonic mode reduces to a classical random variable.  \\

\textbf{Acknowledgments.---} I would like to thank Piper Fowler-Wright for useful discussions on the results of Ref.~\cite{fowler2023}. I am further grateful to Igor Lesanovsky and Albert Cabot for fruitful discussions on related projects. I acknowledge funding from the Deutsche Forschungsgemeinschaft (DFG, German Research Foundation) under Project No. 435696605 and through the Research Unit FOR 5413/1, Grant No. 465199066 as well as from the European Union’s Horizon Europe research and innovation program under Grant Agreement No. 101046968 (BRISQ). I am indebted to the Baden-W\"urttemberg Stiftung for the financial support by the Eliteprogramme for Postdocs.

\bibliography{references}

\begin{thebibliography}{98}%
\makeatletter
\providecommand \@ifxundefined [1]{%
 \@ifx{#1\undefined}
}%
\providecommand \@ifnum [1]{%
 \ifnum #1\expandafter \@firstoftwo
 \else \expandafter \@secondoftwo
 \fi
}%
\providecommand \@ifx [1]{%
 \ifx #1\expandafter \@firstoftwo
 \else \expandafter \@secondoftwo
 \fi
}%
\providecommand \natexlab [1]{#1}%
\providecommand \enquote  [1]{``#1''}%
\providecommand \bibnamefont  [1]{#1}%
\providecommand \bibfnamefont [1]{#1}%
\providecommand \citenamefont [1]{#1}%
\providecommand \href@noop [0]{\@secondoftwo}%
\providecommand \href [0]{\begingroup \@sanitize@url \@href}%
\providecommand \@href[1]{\@@startlink{#1}\@@href}%
\providecommand \@@href[1]{\endgroup#1\@@endlink}%
\providecommand \@sanitize@url [0]{\catcode `\\12\catcode `\$12\catcode
  `\&12\catcode `\#12\catcode `\^12\catcode `\_12\catcode `\%12\relax}%
\providecommand \@@startlink[1]{}%
\providecommand \@@endlink[0]{}%
\providecommand \url  [0]{\begingroup\@sanitize@url \@url }%
\providecommand \@url [1]{\endgroup\@href {#1}{\urlprefix }}%
\providecommand \urlprefix  [0]{URL }%
\providecommand \Eprint [0]{\href }%
\providecommand \doibase [0]{https://doi.org/}%
\providecommand \selectlanguage [0]{\@gobble}%
\providecommand \bibinfo  [0]{\@secondoftwo}%
\providecommand \bibfield  [0]{\@secondoftwo}%
\providecommand \translation [1]{[#1]}%
\providecommand \BibitemOpen [0]{}%
\providecommand \bibitemStop [0]{}%
\providecommand \bibitemNoStop [0]{.\EOS\space}%
\providecommand \EOS [0]{\spacefactor3000\relax}%
\providecommand \BibitemShut  [1]{\csname bibitem#1\endcsname}%
\let\auto@bib@innerbib\@empty
\bibitem [{\citenamefont {Ritsch}\ \emph {et~al.}(2013)\citenamefont {Ritsch},
  \citenamefont {Domokos}, \citenamefont {Brennecke},\ and\ \citenamefont
  {Esslinger}}]{ritsch2013}%
  \BibitemOpen
  \bibfield  {author} {\bibinfo {author} {\bibfnamefont {H.}~\bibnamefont
  {Ritsch}}, \bibinfo {author} {\bibfnamefont {P.}~\bibnamefont {Domokos}},
  \bibinfo {author} {\bibfnamefont {F.}~\bibnamefont {Brennecke}},\ and\
  \bibinfo {author} {\bibfnamefont {T.}~\bibnamefont {Esslinger}},\ }\bibfield
  {title} {\bibinfo {title} {{Cold atoms in cavity-generated dynamical optical
  potentials}},\ }\href {https://doi.org/10.1103/RevModPhys.85.553} {\bibfield
  {journal} {\bibinfo  {journal} {Rev. Mod. Phys.}\ }\textbf {\bibinfo {volume}
  {85}},\ \bibinfo {pages} {553} (\bibinfo {year} {2013})}\BibitemShut
  {NoStop}%
\bibitem [{\citenamefont {Norcia}\ \emph {et~al.}(2018)\citenamefont {Norcia},
  \citenamefont {Lewis-Swan}, \citenamefont {Cline}, \citenamefont {Zhu},
  \citenamefont {Rey},\ and\ \citenamefont {Thompson}}]{norcia2018}%
  \BibitemOpen
  \bibfield  {author} {\bibinfo {author} {\bibfnamefont {M.~A.}\ \bibnamefont
  {Norcia}}, \bibinfo {author} {\bibfnamefont {R.~J.}\ \bibnamefont
  {Lewis-Swan}}, \bibinfo {author} {\bibfnamefont {J.~R.~K.}\ \bibnamefont
  {Cline}}, \bibinfo {author} {\bibfnamefont {B.}~\bibnamefont {Zhu}}, \bibinfo
  {author} {\bibfnamefont {A.~M.}\ \bibnamefont {Rey}},\ and\ \bibinfo {author}
  {\bibfnamefont {J.~K.}\ \bibnamefont {Thompson}},\ }\bibfield  {title}
  {\bibinfo {title} {{Cavity-mediated collective spin-exchange interactions in
  a strontium superradiant laser}},\ }\href
  {https://doi.org/10.1126/science.aar3102} {\bibfield  {journal} {\bibinfo
  {journal} {Science}\ }\textbf {\bibinfo {volume} {361}},\ \bibinfo {pages}
  {259} (\bibinfo {year} {2018})}\BibitemShut {NoStop}%
\bibitem [{\citenamefont {Dogra}\ \emph {et~al.}(2019)\citenamefont {Dogra},
  \citenamefont {Landini}, \citenamefont {Kroeger}, \citenamefont {Hruby},
  \citenamefont {Donner},\ and\ \citenamefont {Esslinger}}]{dogra2019}%
  \BibitemOpen
  \bibfield  {author} {\bibinfo {author} {\bibfnamefont {N.}~\bibnamefont
  {Dogra}}, \bibinfo {author} {\bibfnamefont {M.}~\bibnamefont {Landini}},
  \bibinfo {author} {\bibfnamefont {K.}~\bibnamefont {Kroeger}}, \bibinfo
  {author} {\bibfnamefont {L.}~\bibnamefont {Hruby}}, \bibinfo {author}
  {\bibfnamefont {T.}~\bibnamefont {Donner}},\ and\ \bibinfo {author}
  {\bibfnamefont {T.}~\bibnamefont {Esslinger}},\ }\bibfield  {title} {\bibinfo
  {title} {{Dissipation-induced structural instability and chiral dynamics in a
  quantum gas}},\ }\href {https://doi.org/10.1126/science.aaw4465} {\bibfield
  {journal} {\bibinfo  {journal} {Science}\ }\textbf {\bibinfo {volume}
  {366}},\ \bibinfo {pages} {1496} (\bibinfo {year} {2019})}\BibitemShut
  {NoStop}%
\bibitem [{\citenamefont {Muniz}\ \emph {et~al.}(2020)\citenamefont {Muniz},
  \citenamefont {Barberena}, \citenamefont {Lewis-Swan}, \citenamefont {Young},
  \citenamefont {Cline}, \citenamefont {Rey},\ and\ \citenamefont
  {Thompson}}]{muniz2020}%
  \BibitemOpen
  \bibfield  {author} {\bibinfo {author} {\bibfnamefont {J.~A.}\ \bibnamefont
  {Muniz}}, \bibinfo {author} {\bibfnamefont {D.}~\bibnamefont {Barberena}},
  \bibinfo {author} {\bibfnamefont {R.~J.}\ \bibnamefont {Lewis-Swan}},
  \bibinfo {author} {\bibfnamefont {D.~J.}\ \bibnamefont {Young}}, \bibinfo
  {author} {\bibfnamefont {J.~R.~K.}\ \bibnamefont {Cline}}, \bibinfo {author}
  {\bibfnamefont {A.~M.}\ \bibnamefont {Rey}},\ and\ \bibinfo {author}
  {\bibfnamefont {J.~K.}\ \bibnamefont {Thompson}},\ }\bibfield  {title}
  {\bibinfo {title} {{Exploring dynamical phase transitions with cold atoms in
  an optical cavity}},\ }\href {https://doi.org/10.1038/s41586-020-2224-x}
  {\bibfield  {journal} {\bibinfo  {journal} {Nature}\ }\textbf {\bibinfo
  {volume} {580}},\ \bibinfo {pages} {602} (\bibinfo {year}
  {2020})}\BibitemShut {NoStop}%
\bibitem [{\citenamefont {Mivehvar}\ \emph {et~al.}(2021)\citenamefont
  {Mivehvar}, \citenamefont {Piazza}, \citenamefont {Donner},\ and\
  \citenamefont {Ritsch}}]{mivehvar2021}%
  \BibitemOpen
  \bibfield  {author} {\bibinfo {author} {\bibfnamefont {F.}~\bibnamefont
  {Mivehvar}}, \bibinfo {author} {\bibfnamefont {F.}~\bibnamefont {Piazza}},
  \bibinfo {author} {\bibfnamefont {T.}~\bibnamefont {Donner}},\ and\ \bibinfo
  {author} {\bibfnamefont {H.}~\bibnamefont {Ritsch}},\ }\bibfield  {title}
  {\bibinfo {title} {{Cavity QED with quantum gases: new paradigms in many-body
  physics}},\ }\href {https://doi.org/10.1080/00018732.2021.1969727} {\bibfield
   {journal} {\bibinfo  {journal} {Adv. Phys.}\ }\textbf {\bibinfo {volume}
  {70}},\ \bibinfo {pages} {1} (\bibinfo {year} {2021})}\BibitemShut {NoStop}%
\bibitem [{\citenamefont {Suarez}\ \emph {et~al.}(2023)\citenamefont {Suarez},
  \citenamefont {Carollo}, \citenamefont {Lesanovsky}, \citenamefont {Olmos},
  \citenamefont {Courteille},\ and\ \citenamefont {Slama}}]{suarez2023}%
  \BibitemOpen
  \bibfield  {author} {\bibinfo {author} {\bibfnamefont {E.}~\bibnamefont
  {Suarez}}, \bibinfo {author} {\bibfnamefont {F.}~\bibnamefont {Carollo}},
  \bibinfo {author} {\bibfnamefont {I.}~\bibnamefont {Lesanovsky}}, \bibinfo
  {author} {\bibfnamefont {B.}~\bibnamefont {Olmos}}, \bibinfo {author}
  {\bibfnamefont {P.~W.}\ \bibnamefont {Courteille}},\ and\ \bibinfo {author}
  {\bibfnamefont {S.}~\bibnamefont {Slama}},\ }\bibfield  {title} {\bibinfo
  {title} {{Collective atom-cavity coupling and nonlinear dynamics with atoms
  with multilevel ground states}},\ }\href
  {https://doi.org/10.1103/PhysRevA.107.023714} {\bibfield  {journal} {\bibinfo
   {journal} {Phys. Rev. A}\ }\textbf {\bibinfo {volume} {107}},\ \bibinfo
  {pages} {023714} (\bibinfo {year} {2023})}\BibitemShut {NoStop}%
\bibitem [{\citenamefont {G\'abor}\ \emph {et~al.}(2023)\citenamefont
  {G\'abor}, \citenamefont {Nagy}, \citenamefont {Dombi}, \citenamefont
  {Clark}, \citenamefont {Williams}, \citenamefont {Adwaith}, \citenamefont
  {Vukics},\ and\ \citenamefont {Domokos}}]{gabor2023}%
  \BibitemOpen
  \bibfield  {author} {\bibinfo {author} {\bibfnamefont {B.}~\bibnamefont
  {G\'abor}}, \bibinfo {author} {\bibfnamefont {D.}~\bibnamefont {Nagy}},
  \bibinfo {author} {\bibfnamefont {A.}~\bibnamefont {Dombi}}, \bibinfo
  {author} {\bibfnamefont {T.~W.}\ \bibnamefont {Clark}}, \bibinfo {author}
  {\bibfnamefont {F.~I.~B.}\ \bibnamefont {Williams}}, \bibinfo {author}
  {\bibfnamefont {K.~V.}\ \bibnamefont {Adwaith}}, \bibinfo {author}
  {\bibfnamefont {A.}~\bibnamefont {Vukics}},\ and\ \bibinfo {author}
  {\bibfnamefont {P.}~\bibnamefont {Domokos}},\ }\bibfield  {title} {\bibinfo
  {title} {{Ground-state bistability of cold atoms in a cavity}},\ }\href
  {https://doi.org/10.1103/PhysRevA.107.023713} {\bibfield  {journal} {\bibinfo
   {journal} {Phys. Rev. A}\ }\textbf {\bibinfo {volume} {107}},\ \bibinfo
  {pages} {023713} (\bibinfo {year} {2023})}\BibitemShut {NoStop}%
\bibitem [{\citenamefont {Hepp}\ and\ \citenamefont
  {Lieb}(1973{\natexlab{a}})}]{hepp1973}%
  \BibitemOpen
  \bibfield  {author} {\bibinfo {author} {\bibfnamefont {K.}~\bibnamefont
  {Hepp}}\ and\ \bibinfo {author} {\bibfnamefont {E.~H.}\ \bibnamefont
  {Lieb}},\ }\bibfield  {title} {\bibinfo {title} {{On the superradiant phase
  transition for molecules in a quantized radiation field: the Dicke maser
  model}},\ }\href
  {https://doi.org/https://doi.org/10.1016/0003-4916(73)90039-0} {\bibfield
  {journal} {\bibinfo  {journal} {Ann. Phys.}\ }\textbf {\bibinfo {volume}
  {76}},\ \bibinfo {pages} {360} (\bibinfo {year}
  {1973}{\natexlab{a}})}\BibitemShut {NoStop}%
\bibitem [{\citenamefont {Hepp}\ and\ \citenamefont
  {Lieb}(1973{\natexlab{b}})}]{hepp1973b}%
  \BibitemOpen
  \bibfield  {author} {\bibinfo {author} {\bibfnamefont {K.}~\bibnamefont
  {Hepp}}\ and\ \bibinfo {author} {\bibfnamefont {E.~H.}\ \bibnamefont
  {Lieb}},\ }\bibfield  {title} {\bibinfo {title} {{Equilibrium Statistical
  Mechanics of Matter Interacting with the Quantized Radiation Field}},\ }\href
  {https://doi.org/10.1103/PhysRevA.8.2517} {\bibfield  {journal} {\bibinfo
  {journal} {Phys. Rev. A}\ }\textbf {\bibinfo {volume} {8}},\ \bibinfo {pages}
  {2517} (\bibinfo {year} {1973}{\natexlab{b}})}\BibitemShut {NoStop}%
\bibitem [{\citenamefont {Emary}\ and\ \citenamefont
  {Brandes}(2003{\natexlab{a}})}]{emary2003}%
  \BibitemOpen
  \bibfield  {author} {\bibinfo {author} {\bibfnamefont {C.}~\bibnamefont
  {Emary}}\ and\ \bibinfo {author} {\bibfnamefont {T.}~\bibnamefont
  {Brandes}},\ }\bibfield  {title} {\bibinfo {title} {{Quantum Chaos Triggered
  by Precursors of a Quantum Phase Transition: The Dicke Model}},\ }\href
  {https://doi.org/10.1103/PhysRevLett.90.044101} {\bibfield  {journal}
  {\bibinfo  {journal} {Phys. Rev. Lett.}\ }\textbf {\bibinfo {volume} {90}},\
  \bibinfo {pages} {044101} (\bibinfo {year} {2003}{\natexlab{a}})}\BibitemShut
  {NoStop}%
\bibitem [{\citenamefont {Emary}\ and\ \citenamefont
  {Brandes}(2003{\natexlab{b}})}]{emary2003b}%
  \BibitemOpen
  \bibfield  {author} {\bibinfo {author} {\bibfnamefont {C.}~\bibnamefont
  {Emary}}\ and\ \bibinfo {author} {\bibfnamefont {T.}~\bibnamefont
  {Brandes}},\ }\bibfield  {title} {\bibinfo {title} {{Chaos and the quantum
  phase transition in the Dicke model}},\ }\href
  {https://doi.org/10.1103/PhysRevE.67.066203} {\bibfield  {journal} {\bibinfo
  {journal} {Phys. Rev. E}\ }\textbf {\bibinfo {volume} {67}},\ \bibinfo
  {pages} {066203} (\bibinfo {year} {2003}{\natexlab{b}})}\BibitemShut
  {NoStop}%
\bibitem [{\citenamefont {Chase}\ and\ \citenamefont
  {Geremia}(2008)}]{chase2008}%
  \BibitemOpen
  \bibfield  {author} {\bibinfo {author} {\bibfnamefont {B.~A.}\ \bibnamefont
  {Chase}}\ and\ \bibinfo {author} {\bibfnamefont {J.~M.}\ \bibnamefont
  {Geremia}},\ }\bibfield  {title} {\bibinfo {title} {{Collective processes of
  an ensemble of spin-$1/2$ particles}},\ }\href
  {https://doi.org/10.1103/PhysRevA.78.052101} {\bibfield  {journal} {\bibinfo
  {journal} {Phys. Rev. A}\ }\textbf {\bibinfo {volume} {78}},\ \bibinfo
  {pages} {052101} (\bibinfo {year} {2008})}\BibitemShut {NoStop}%
\bibitem [{\citenamefont {Baragiola}\ \emph {et~al.}(2010)\citenamefont
  {Baragiola}, \citenamefont {Chase},\ and\ \citenamefont
  {Geremia}}]{baragiola2010}%
  \BibitemOpen
  \bibfield  {author} {\bibinfo {author} {\bibfnamefont {B.~Q.}\ \bibnamefont
  {Baragiola}}, \bibinfo {author} {\bibfnamefont {B.~A.}\ \bibnamefont
  {Chase}},\ and\ \bibinfo {author} {\bibfnamefont {J.}~\bibnamefont
  {Geremia}},\ }\bibfield  {title} {\bibinfo {title} {{Collective uncertainty
  in partially polarized and partially decohered spin-$\frac{1}{2}$ systems}},\
  }\href {https://doi.org/10.1103/PhysRevA.81.032104} {\bibfield  {journal}
  {\bibinfo  {journal} {Phys. Rev. A}\ }\textbf {\bibinfo {volume} {81}},\
  \bibinfo {pages} {032104} (\bibinfo {year} {2010})}\BibitemShut {NoStop}%
\bibitem [{\citenamefont {Sieberer}\ \emph {et~al.}(2016)\citenamefont
  {Sieberer}, \citenamefont {Buchhold},\ and\ \citenamefont
  {Diehl}}]{sieberer2016}%
  \BibitemOpen
  \bibfield  {author} {\bibinfo {author} {\bibfnamefont {L.~M.}\ \bibnamefont
  {Sieberer}}, \bibinfo {author} {\bibfnamefont {M.}~\bibnamefont {Buchhold}},\
  and\ \bibinfo {author} {\bibfnamefont {S.}~\bibnamefont {Diehl}},\ }\bibfield
   {title} {\bibinfo {title} {Keldysh field theory for driven open quantum
  systems},\ }\href {https://doi.org/10.1088/0034-4885/79/9/096001} {\bibfield
  {journal} {\bibinfo  {journal} {Rep. Prog. Phys.}\ }\textbf {\bibinfo
  {volume} {79}},\ \bibinfo {pages} {096001} (\bibinfo {year}
  {2016})}\BibitemShut {NoStop}%
\bibitem [{\citenamefont {Kirton}\ and\ \citenamefont
  {Keeling}(2017)}]{kirton2017}%
  \BibitemOpen
  \bibfield  {author} {\bibinfo {author} {\bibfnamefont {P.}~\bibnamefont
  {Kirton}}\ and\ \bibinfo {author} {\bibfnamefont {J.}~\bibnamefont
  {Keeling}},\ }\bibfield  {title} {\bibinfo {title} {{Suppressing and
  Restoring the Dicke Superradiance Transition by Dephasing and Decay}},\
  }\href {https://doi.org/10.1103/PhysRevLett.118.123602} {\bibfield  {journal}
  {\bibinfo  {journal} {Phys. Rev. Lett.}\ }\textbf {\bibinfo {volume} {118}},\
  \bibinfo {pages} {123602} (\bibinfo {year} {2017})}\BibitemShut {NoStop}%
\bibitem [{\citenamefont {Shammah}\ \emph {et~al.}(2018)\citenamefont
  {Shammah}, \citenamefont {Ahmed}, \citenamefont {Lambert}, \citenamefont
  {De~Liberato},\ and\ \citenamefont {Nori}}]{shammah2018}%
  \BibitemOpen
  \bibfield  {author} {\bibinfo {author} {\bibfnamefont {N.}~\bibnamefont
  {Shammah}}, \bibinfo {author} {\bibfnamefont {S.}~\bibnamefont {Ahmed}},
  \bibinfo {author} {\bibfnamefont {N.}~\bibnamefont {Lambert}}, \bibinfo
  {author} {\bibfnamefont {S.}~\bibnamefont {De~Liberato}},\ and\ \bibinfo
  {author} {\bibfnamefont {F.}~\bibnamefont {Nori}},\ }\bibfield  {title}
  {\bibinfo {title} {{Open quantum systems with local and collective incoherent
  processes: Efficient numerical simulations using permutational invariance}},\
  }\href {https://doi.org/10.1103/PhysRevA.98.063815} {\bibfield  {journal}
  {\bibinfo  {journal} {Phys. Rev. A}\ }\textbf {\bibinfo {volume} {98}},\
  \bibinfo {pages} {063815} (\bibinfo {year} {2018})}\BibitemShut {NoStop}%
\bibitem [{\citenamefont {Reitz}\ \emph {et~al.}(2022)\citenamefont {Reitz},
  \citenamefont {Sommer},\ and\ \citenamefont {Genes}}]{reitz2022}%
  \BibitemOpen
  \bibfield  {author} {\bibinfo {author} {\bibfnamefont {M.}~\bibnamefont
  {Reitz}}, \bibinfo {author} {\bibfnamefont {C.}~\bibnamefont {Sommer}},\ and\
  \bibinfo {author} {\bibfnamefont {C.}~\bibnamefont {Genes}},\ }\bibfield
  {title} {\bibinfo {title} {{Cooperative Quantum Phenomena in Light-Matter
  Platforms}},\ }\href {https://doi.org/10.1103/PRXQuantum.3.010201} {\bibfield
   {journal} {\bibinfo  {journal} {PRX Quantum}\ }\textbf {\bibinfo {volume}
  {3}},\ \bibinfo {pages} {010201} (\bibinfo {year} {2022})}\BibitemShut
  {NoStop}%
\bibitem [{\citenamefont {Spohn}(1980)}]{spohn1980}%
  \BibitemOpen
  \bibfield  {author} {\bibinfo {author} {\bibfnamefont {H.}~\bibnamefont
  {Spohn}},\ }\bibfield  {title} {\bibinfo {title} {{Kinetic equations from
  Hamiltonian dynamics: Markovian limits}},\ }\href
  {https://doi.org/10.1103/RevModPhys.52.569} {\bibfield  {journal} {\bibinfo
  {journal} {Rev. Mod. Phys.}\ }\textbf {\bibinfo {volume} {52}},\ \bibinfo
  {pages} {569} (\bibinfo {year} {1980})}\BibitemShut {NoStop}%
\bibitem [{\citenamefont {Alicki}\ and\ \citenamefont
  {Messer}(1983)}]{alicki1983}%
  \BibitemOpen
  \bibfield  {author} {\bibinfo {author} {\bibfnamefont {R.}~\bibnamefont
  {Alicki}}\ and\ \bibinfo {author} {\bibfnamefont {J.}~\bibnamefont
  {Messer}},\ }\bibfield  {title} {\bibinfo {title} {{Nonlinear quantum
  dynamical semigroups for many-body open systems}},\ }\href
  {https://doi.org/10.1007/BF01012712} {\bibfield  {journal} {\bibinfo
  {journal} {J. Stat. Phys.}\ }\textbf {\bibinfo {volume} {32}},\ \bibinfo
  {pages} {299} (\bibinfo {year} {1983})}\BibitemShut {NoStop}%
\bibitem [{\citenamefont {Mori}(2013)}]{mori2013}%
  \BibitemOpen
  \bibfield  {author} {\bibinfo {author} {\bibfnamefont {T.}~\bibnamefont
  {Mori}},\ }\bibfield  {title} {\bibinfo {title} {{Exactness of the mean-field
  dynamics in optical cavity systems}},\ }\href
  {https://doi.org/10.1088/1742-5468/2013/06/P06005} {\bibfield  {journal}
  {\bibinfo  {journal} {J. Stat. Mech. Theory Exp.}\ }\textbf {\bibinfo
  {volume} {2013}},\ \bibinfo {pages} {P06005} (\bibinfo {year}
  {2013})}\BibitemShut {NoStop}%
\bibitem [{\citenamefont {Benatti}\ \emph {et~al.}(2016)\citenamefont
  {Benatti}, \citenamefont {Carollo}, \citenamefont {Floreanini},\ and\
  \citenamefont {Narnhofer}}]{benatti2016}%
  \BibitemOpen
  \bibfield  {author} {\bibinfo {author} {\bibfnamefont {F.}~\bibnamefont
  {Benatti}}, \bibinfo {author} {\bibfnamefont {F.}~\bibnamefont {Carollo}},
  \bibinfo {author} {\bibfnamefont {R.}~\bibnamefont {Floreanini}},\ and\
  \bibinfo {author} {\bibfnamefont {H.}~\bibnamefont {Narnhofer}},\ }\bibfield
  {title} {\bibinfo {title} {{Non-markovian mesoscopic dissipative dynamics of
  open quantum spin chains}},\ }\href
  {https://doi.org/https://doi.org/10.1016/j.physleta.2015.10.062} {\bibfield
  {journal} {\bibinfo  {journal} {Phys. Lett. A}\ }\textbf {\bibinfo {volume}
  {380}},\ \bibinfo {pages} {381} (\bibinfo {year} {2016})}\BibitemShut
  {NoStop}%
\bibitem [{\citenamefont {Merkli}\ and\ \citenamefont
  {Rafiyi}(2018)}]{merkli2018}%
  \BibitemOpen
  \bibfield  {author} {\bibinfo {author} {\bibfnamefont {M.}~\bibnamefont
  {Merkli}}\ and\ \bibinfo {author} {\bibfnamefont {A.}~\bibnamefont
  {Rafiyi}},\ }\bibfield  {title} {\bibinfo {title} {{Mean field dynamics of
  some open quantum systems}},\ }\href {https://doi.org/10.1098/rspa.2017.0856}
  {\bibfield  {journal} {\bibinfo  {journal} {Proc. R. Soc. A: Math.}\ }\textbf
  {\bibinfo {volume} {474}},\ \bibinfo {pages} {20170856} (\bibinfo {year}
  {2018})}\BibitemShut {NoStop}%
\bibitem [{\citenamefont {Benatti}\ \emph {et~al.}(2018)\citenamefont
  {Benatti}, \citenamefont {Carollo}, \citenamefont {Floreanini},\ and\
  \citenamefont {Narnhofer}}]{benatti2018}%
  \BibitemOpen
  \bibfield  {author} {\bibinfo {author} {\bibfnamefont {F.}~\bibnamefont
  {Benatti}}, \bibinfo {author} {\bibfnamefont {F.}~\bibnamefont {Carollo}},
  \bibinfo {author} {\bibfnamefont {R.}~\bibnamefont {Floreanini}},\ and\
  \bibinfo {author} {\bibfnamefont {H.}~\bibnamefont {Narnhofer}},\ }\bibfield
  {title} {\bibinfo {title} {{Quantum spin chain dissipative mean-field
  dynamics}},\ }\href {https://dx.doi.org/10.1088/1751-8121/aacbdb} {\bibfield
  {journal} {\bibinfo  {journal} {J. Phys. A}\ }\textbf {\bibinfo {volume}
  {51}},\ \bibinfo {pages} {325001} (\bibinfo {year} {2018})}\BibitemShut
  {NoStop}%
\bibitem [{\citenamefont {Carollo}\ and\ \citenamefont
  {Lesanovsky}(2021)}]{carollo2021}%
  \BibitemOpen
  \bibfield  {author} {\bibinfo {author} {\bibfnamefont {F.}~\bibnamefont
  {Carollo}}\ and\ \bibinfo {author} {\bibfnamefont {I.}~\bibnamefont
  {Lesanovsky}},\ }\bibfield  {title} {\bibinfo {title} {{Exactness of
  Mean-Field Equations for Open Dicke Models with an Application to Pattern
  Retrieval Dynamics}},\ }\href
  {https://doi.org/10.1103/PhysRevLett.126.230601} {\bibfield  {journal}
  {\bibinfo  {journal} {Phys. Rev. Lett.}\ }\textbf {\bibinfo {volume} {126}},\
  \bibinfo {pages} {230601} (\bibinfo {year} {2021})}\BibitemShut {NoStop}%
\bibitem [{\citenamefont {Dicke}(1954)}]{dicke1954}%
  \BibitemOpen
  \bibfield  {author} {\bibinfo {author} {\bibfnamefont {R.~H.}\ \bibnamefont
  {Dicke}},\ }\bibfield  {title} {\bibinfo {title} {{Coherence in Spontaneous
  Radiation Processes}},\ }\href {https://doi.org/10.1103/PhysRev.93.99}
  {\bibfield  {journal} {\bibinfo  {journal} {Phys. Rev.}\ }\textbf {\bibinfo
  {volume} {93}},\ \bibinfo {pages} {99} (\bibinfo {year} {1954})}\BibitemShut
  {NoStop}%
\bibitem [{\citenamefont {Wang}\ and\ \citenamefont {Hioe}(1973)}]{wang1973}%
  \BibitemOpen
  \bibfield  {author} {\bibinfo {author} {\bibfnamefont {Y.~K.}\ \bibnamefont
  {Wang}}\ and\ \bibinfo {author} {\bibfnamefont {F.~T.}\ \bibnamefont
  {Hioe}},\ }\bibfield  {title} {\bibinfo {title} {{Phase Transition in the
  Dicke Model of Superradiance}},\ }\href
  {https://doi.org/10.1103/PhysRevA.7.831} {\bibfield  {journal} {\bibinfo
  {journal} {Phys. Rev. A}\ }\textbf {\bibinfo {volume} {7}},\ \bibinfo {pages}
  {831} (\bibinfo {year} {1973})}\BibitemShut {NoStop}%
\bibitem [{\citenamefont {Hioe}(1973)}]{hioe1973}%
  \BibitemOpen
  \bibfield  {author} {\bibinfo {author} {\bibfnamefont {F.~T.}\ \bibnamefont
  {Hioe}},\ }\bibfield  {title} {\bibinfo {title} {{Phase Transitions in Some
  Generalized Dicke Models of Superradiance}},\ }\href
  {https://doi.org/10.1103/PhysRevA.8.1440} {\bibfield  {journal} {\bibinfo
  {journal} {Phys. Rev. A}\ }\textbf {\bibinfo {volume} {8}},\ \bibinfo {pages}
  {1440} (\bibinfo {year} {1973})}\BibitemShut {NoStop}%
\bibitem [{\citenamefont {Carmichael}\ \emph {et~al.}(1973)\citenamefont
  {Carmichael}, \citenamefont {Gardiner},\ and\ \citenamefont
  {Walls}}]{carmichael1973}%
  \BibitemOpen
  \bibfield  {author} {\bibinfo {author} {\bibfnamefont {H.}~\bibnamefont
  {Carmichael}}, \bibinfo {author} {\bibfnamefont {C.}~\bibnamefont
  {Gardiner}},\ and\ \bibinfo {author} {\bibfnamefont {D.}~\bibnamefont
  {Walls}},\ }\bibfield  {title} {\bibinfo {title} {{Higher order corrections
  to the Dicke superradiant phase transition}},\ }\href
  {https://doi.org/https://doi.org/10.1016/0375-9601(73)90679-8} {\bibfield
  {journal} {\bibinfo  {journal} {Phys. Lett. A}\ }\textbf {\bibinfo {volume}
  {46}},\ \bibinfo {pages} {47} (\bibinfo {year} {1973})}\BibitemShut {NoStop}%
\bibitem [{\citenamefont {S\'anchez Mu\~noz}\ \emph {et~al.}(2019)\citenamefont
  {S\'anchez Mu\~noz}, \citenamefont {Bu\ifmmode~\check{c}\else \v{c}\fi{}a},
  \citenamefont {Tindall}, \citenamefont {Gonz\'alez-Tudela}, \citenamefont
  {Jaksch},\ and\ \citenamefont {Porras}}]{sanchez2019}%
  \BibitemOpen
  \bibfield  {author} {\bibinfo {author} {\bibfnamefont {C.}~\bibnamefont
  {S\'anchez Mu\~noz}}, \bibinfo {author} {\bibfnamefont {B.}~\bibnamefont
  {Bu\ifmmode~\check{c}\else \v{c}\fi{}a}}, \bibinfo {author} {\bibfnamefont
  {J.}~\bibnamefont {Tindall}}, \bibinfo {author} {\bibfnamefont
  {A.}~\bibnamefont {Gonz\'alez-Tudela}}, \bibinfo {author} {\bibfnamefont
  {D.}~\bibnamefont {Jaksch}},\ and\ \bibinfo {author} {\bibfnamefont
  {D.}~\bibnamefont {Porras}},\ }\bibfield  {title} {\bibinfo {title}
  {{Symmetries and conservation laws in quantum trajectories: Dissipative
  freezing}},\ }\href {https://doi.org/10.1103/PhysRevA.100.042113} {\bibfield
  {journal} {\bibinfo  {journal} {Phys. Rev. A}\ }\textbf {\bibinfo {volume}
  {100}},\ \bibinfo {pages} {042113} (\bibinfo {year} {2019})}\BibitemShut
  {NoStop}%
\bibitem [{\citenamefont {Bu\ifmmode~\check{c}\else \v{c}\fi{}a}\ and\
  \citenamefont {Jaksch}(2019)}]{buca2019}%
  \BibitemOpen
  \bibfield  {author} {\bibinfo {author} {\bibfnamefont {B.}~\bibnamefont
  {Bu\ifmmode~\check{c}\else \v{c}\fi{}a}}\ and\ \bibinfo {author}
  {\bibfnamefont {D.}~\bibnamefont {Jaksch}},\ }\bibfield  {title} {\bibinfo
  {title} {{Dissipation Induced Nonstationarity in a Quantum Gas}},\ }\href
  {https://doi.org/10.1103/PhysRevLett.123.260401} {\bibfield  {journal}
  {\bibinfo  {journal} {Phys. Rev. Lett.}\ }\textbf {\bibinfo {volume} {123}},\
  \bibinfo {pages} {260401} (\bibinfo {year} {2019})}\BibitemShut {NoStop}%
\bibitem [{\citenamefont {Tomadin}\ and\ \citenamefont
  {Fazio}(2010)}]{tomadin2010}%
  \BibitemOpen
  \bibfield  {author} {\bibinfo {author} {\bibfnamefont {A.}~\bibnamefont
  {Tomadin}}\ and\ \bibinfo {author} {\bibfnamefont {R.}~\bibnamefont
  {Fazio}},\ }\bibfield  {title} {\bibinfo {title} {{Many-body phenomena in
  QED-cavity arrays}},\ }\href {https://doi.org/10.1364/JOSAB.27.00A130}
  {\bibfield  {journal} {\bibinfo  {journal} {J. Opt. Soc. Am. B}\ }\textbf
  {\bibinfo {volume} {27}},\ \bibinfo {pages} {A130} (\bibinfo {year}
  {2010})}\BibitemShut {NoStop}%
\bibitem [{\citenamefont {Kirton}\ \emph {et~al.}(2019)\citenamefont {Kirton},
  \citenamefont {Roses}, \citenamefont {Keeling},\ and\ \citenamefont
  {Dalla~Torre}}]{kirton2019}%
  \BibitemOpen
  \bibfield  {author} {\bibinfo {author} {\bibfnamefont {P.}~\bibnamefont
  {Kirton}}, \bibinfo {author} {\bibfnamefont {M.~M.}\ \bibnamefont {Roses}},
  \bibinfo {author} {\bibfnamefont {J.}~\bibnamefont {Keeling}},\ and\ \bibinfo
  {author} {\bibfnamefont {E.~G.}\ \bibnamefont {Dalla~Torre}},\ }\bibfield
  {title} {\bibinfo {title} {{Introduction to the Dicke Model: From Equilibrium
  to Nonequilibrium, and Vice Versa}},\ }\href
  {https://doi.org/https://doi.org/10.1002/qute.201800043} {\bibfield
  {journal} {\bibinfo  {journal} {Adv. Quantum Technol.}\ }\textbf {\bibinfo
  {volume} {2}},\ \bibinfo {pages} {1800043} (\bibinfo {year}
  {2019})}\BibitemShut {NoStop}%
\bibitem [{\citenamefont {Boneberg}\ \emph {et~al.}(2022)\citenamefont
  {Boneberg}, \citenamefont {Lesanovsky},\ and\ \citenamefont
  {Carollo}}]{boneberg2022}%
  \BibitemOpen
  \bibfield  {author} {\bibinfo {author} {\bibfnamefont {M.}~\bibnamefont
  {Boneberg}}, \bibinfo {author} {\bibfnamefont {I.}~\bibnamefont
  {Lesanovsky}},\ and\ \bibinfo {author} {\bibfnamefont {F.}~\bibnamefont
  {Carollo}},\ }\bibfield  {title} {\bibinfo {title} {{Quantum fluctuations and
  correlations in open quantum Dicke models}},\ }\href
  {https://doi.org/10.1103/PhysRevA.106.012212} {\bibfield  {journal} {\bibinfo
   {journal} {Phys. Rev. A}\ }\textbf {\bibinfo {volume} {106}},\ \bibinfo
  {pages} {012212} (\bibinfo {year} {2022})}\BibitemShut {NoStop}%
\bibitem [{\citenamefont {Mattes}\ \emph {et~al.}(2023)\citenamefont {Mattes},
  \citenamefont {Lesanovsky},\ and\ \citenamefont {Carollo}}]{mattes2023}%
  \BibitemOpen
  \bibfield  {author} {\bibinfo {author} {\bibfnamefont {R.}~\bibnamefont
  {Mattes}}, \bibinfo {author} {\bibfnamefont {I.}~\bibnamefont {Lesanovsky}},\
  and\ \bibinfo {author} {\bibfnamefont {F.}~\bibnamefont {Carollo}},\
  }\bibfield  {title} {\bibinfo {title} {Entangled time-crystal phase in an
  open quantum light-matter system},\ }\href
  {https://doi.org/10.48550/arXiv.2303.07725} {\bibfield  {journal} {\bibinfo
  {journal} {arXiv:2303.07725}\ } (\bibinfo {year} {2023})}\BibitemShut
  {NoStop}%
\bibitem [{\citenamefont {Fiorelli}\ \emph {et~al.}(2023)\citenamefont
  {Fiorelli}, \citenamefont {M\"uller}, \citenamefont {Lesanovsky},\ and\
  \citenamefont {Carollo}}]{fiorelli2023}%
  \BibitemOpen
  \bibfield  {author} {\bibinfo {author} {\bibfnamefont {E.}~\bibnamefont
  {Fiorelli}}, \bibinfo {author} {\bibfnamefont {M.}~\bibnamefont {M\"uller}},
  \bibinfo {author} {\bibfnamefont {I.}~\bibnamefont {Lesanovsky}},\ and\
  \bibinfo {author} {\bibfnamefont {F.}~\bibnamefont {Carollo}},\ }\bibfield
  {title} {\bibinfo {title} {{Mean-field dynamics of open quantum systems with
  collective operator-valued rates: validity and application}},\ }\href
  {https://doi.org/10.48550/ARXIV.2302.04155} {\bibfield  {journal} {\bibinfo
  {journal} {arXiv:2302.04155}\ } (\bibinfo {year} {2023})}\BibitemShut
  {NoStop}%
\bibitem [{\citenamefont {Yuzbashyan}\ \emph {et~al.}(2005)\citenamefont
  {Yuzbashyan}, \citenamefont {Altshuler}, \citenamefont {Kuznetsov},\ and\
  \citenamefont {Enolskii}}]{yuzbashyan2005}%
  \BibitemOpen
  \bibfield  {author} {\bibinfo {author} {\bibfnamefont {E.~A.}\ \bibnamefont
  {Yuzbashyan}}, \bibinfo {author} {\bibfnamefont {B.~L.}\ \bibnamefont
  {Altshuler}}, \bibinfo {author} {\bibfnamefont {V.~B.}\ \bibnamefont
  {Kuznetsov}},\ and\ \bibinfo {author} {\bibfnamefont {V.~Z.}\ \bibnamefont
  {Enolskii}},\ }\bibfield  {title} {\bibinfo {title} {{Solution for the
  dynamics of the BCS and central spin problems}},\ }\href
  {https://doi.org/10.1088/0305-4470/38/36/003} {\bibfield  {journal} {\bibinfo
   {journal} {J. Phys. A: Math. Gen.}\ }\textbf {\bibinfo {volume} {38}},\
  \bibinfo {pages} {7831} (\bibinfo {year} {2005})}\BibitemShut {NoStop}%
\bibitem [{\citenamefont {Bortz}\ and\ \citenamefont
  {Stolze}(2007{\natexlab{a}})}]{bortz2007}%
  \BibitemOpen
  \bibfield  {author} {\bibinfo {author} {\bibfnamefont {M.}~\bibnamefont
  {Bortz}}\ and\ \bibinfo {author} {\bibfnamefont {J.}~\bibnamefont {Stolze}},\
  }\bibfield  {title} {\bibinfo {title} {Spin and entanglement dynamics in the
  central-spin model with homogeneous couplings},\ }\href
  {https://doi.org/10.1088/1742-5468/2007/06/P06018} {\bibfield  {journal}
  {\bibinfo  {journal} {J. Stat. Mech.}\ }\textbf {\bibinfo {volume} {2007}},\
  \bibinfo {pages} {P06018} (\bibinfo {year} {2007}{\natexlab{a}})}\BibitemShut
  {NoStop}%
\bibitem [{\citenamefont {Bortz}\ and\ \citenamefont
  {Stolze}(2007{\natexlab{b}})}]{bortz2007b}%
  \BibitemOpen
  \bibfield  {author} {\bibinfo {author} {\bibfnamefont {M.}~\bibnamefont
  {Bortz}}\ and\ \bibinfo {author} {\bibfnamefont {J.}~\bibnamefont {Stolze}},\
  }\bibfield  {title} {\bibinfo {title} {{Exact dynamics in the inhomogeneous
  central-spin model}},\ }\href {https://doi.org/10.1103/PhysRevB.76.014304}
  {\bibfield  {journal} {\bibinfo  {journal} {Phys. Rev. B}\ }\textbf {\bibinfo
  {volume} {76}},\ \bibinfo {pages} {014304} (\bibinfo {year}
  {2007}{\natexlab{b}})}\BibitemShut {NoStop}%
\bibitem [{\citenamefont {Coish}\ \emph {et~al.}(2007)\citenamefont {Coish},
  \citenamefont {Loss}, \citenamefont {Yuzbashyan},\ and\ \citenamefont
  {Altshuler}}]{coish2007}%
  \BibitemOpen
  \bibfield  {author} {\bibinfo {author} {\bibfnamefont {W.~A.}\ \bibnamefont
  {Coish}}, \bibinfo {author} {\bibfnamefont {D.}~\bibnamefont {Loss}},
  \bibinfo {author} {\bibfnamefont {E.~A.}\ \bibnamefont {Yuzbashyan}},\ and\
  \bibinfo {author} {\bibfnamefont {B.~L.}\ \bibnamefont {Altshuler}},\
  }\bibfield  {title} {\bibinfo {title} {Quantum versus classical
  hyperfine-induced dynamics in a quantum dot},\ }\href
  {https://doi.org/10.1063/1.2722783} {\bibfield  {journal} {\bibinfo
  {journal} {J. Appl. Phys.}\ }\textbf {\bibinfo {volume} {101}},\ \bibinfo
  {pages} {081715} (\bibinfo {year} {2007})}\BibitemShut {NoStop}%
\bibitem [{\citenamefont {Maletinsky}\ \emph {et~al.}(2009)\citenamefont
  {Maletinsky}, \citenamefont {Kroner},\ and\ \citenamefont
  {Imamoglu}}]{maletinsky2009}%
  \BibitemOpen
  \bibfield  {author} {\bibinfo {author} {\bibfnamefont {P.}~\bibnamefont
  {Maletinsky}}, \bibinfo {author} {\bibfnamefont {M.}~\bibnamefont {Kroner}},\
  and\ \bibinfo {author} {\bibfnamefont {A.}~\bibnamefont {Imamoglu}},\
  }\bibfield  {title} {\bibinfo {title} {Breakdown of the
  nuclear-spin-temperature approach in quantum-dot demagnetization
  experiments},\ }\href {https://doi.org/10.1038/nphys1273} {\bibfield
  {journal} {\bibinfo  {journal} {Nat. Phys.}\ }\textbf {\bibinfo {volume}
  {5}},\ \bibinfo {pages} {407} (\bibinfo {year} {2009})}\BibitemShut {NoStop}%
\bibitem [{\citenamefont {Kessler}\ \emph {et~al.}(2010)\citenamefont
  {Kessler}, \citenamefont {Yelin}, \citenamefont {Lukin}, \citenamefont
  {Cirac},\ and\ \citenamefont {Giedke}}]{kessler2010}%
  \BibitemOpen
  \bibfield  {author} {\bibinfo {author} {\bibfnamefont {E.~M.}\ \bibnamefont
  {Kessler}}, \bibinfo {author} {\bibfnamefont {S.}~\bibnamefont {Yelin}},
  \bibinfo {author} {\bibfnamefont {M.~D.}\ \bibnamefont {Lukin}}, \bibinfo
  {author} {\bibfnamefont {J.~I.}\ \bibnamefont {Cirac}},\ and\ \bibinfo
  {author} {\bibfnamefont {G.}~\bibnamefont {Giedke}},\ }\bibfield  {title}
  {\bibinfo {title} {{Optical Superradiance from Nuclear Spin Environment of
  Single-Photon Emitters}},\ }\href
  {https://doi.org/10.1103/PhysRevLett.104.143601} {\bibfield  {journal}
  {\bibinfo  {journal} {Phys. Rev. Lett.}\ }\textbf {\bibinfo {volume} {104}},\
  \bibinfo {pages} {143601} (\bibinfo {year} {2010})}\BibitemShut {NoStop}%
\bibitem [{\citenamefont {Kessler}\ \emph {et~al.}(2012)\citenamefont
  {Kessler}, \citenamefont {Giedke}, \citenamefont {Imamoglu}, \citenamefont
  {Yelin}, \citenamefont {Lukin},\ and\ \citenamefont {Cirac}}]{kessler2012}%
  \BibitemOpen
  \bibfield  {author} {\bibinfo {author} {\bibfnamefont {E.~M.}\ \bibnamefont
  {Kessler}}, \bibinfo {author} {\bibfnamefont {G.}~\bibnamefont {Giedke}},
  \bibinfo {author} {\bibfnamefont {A.}~\bibnamefont {Imamoglu}}, \bibinfo
  {author} {\bibfnamefont {S.~F.}\ \bibnamefont {Yelin}}, \bibinfo {author}
  {\bibfnamefont {M.~D.}\ \bibnamefont {Lukin}},\ and\ \bibinfo {author}
  {\bibfnamefont {J.~I.}\ \bibnamefont {Cirac}},\ }\bibfield  {title} {\bibinfo
  {title} {{Dissipative phase transition in a central spin system}},\ }\href
  {https://doi.org/10.1103/PhysRevA.86.012116} {\bibfield  {journal} {\bibinfo
  {journal} {Phys. Rev. A}\ }\textbf {\bibinfo {volume} {86}},\ \bibinfo
  {pages} {012116} (\bibinfo {year} {2012})}\BibitemShut {NoStop}%
\bibitem [{\citenamefont {Schwartz}\ \emph {et~al.}(2016)\citenamefont
  {Schwartz}, \citenamefont {Cogan}, \citenamefont {Schmidgall}, \citenamefont
  {Don}, \citenamefont {Gantz}, \citenamefont {Kenneth}, \citenamefont
  {Lindner},\ and\ \citenamefont {Gershoni}}]{schwartz2016}%
  \BibitemOpen
  \bibfield  {author} {\bibinfo {author} {\bibfnamefont {I.}~\bibnamefont
  {Schwartz}}, \bibinfo {author} {\bibfnamefont {D.}~\bibnamefont {Cogan}},
  \bibinfo {author} {\bibfnamefont {E.~R.}\ \bibnamefont {Schmidgall}},
  \bibinfo {author} {\bibfnamefont {Y.}~\bibnamefont {Don}}, \bibinfo {author}
  {\bibfnamefont {L.}~\bibnamefont {Gantz}}, \bibinfo {author} {\bibfnamefont
  {O.}~\bibnamefont {Kenneth}}, \bibinfo {author} {\bibfnamefont {N.~H.}\
  \bibnamefont {Lindner}},\ and\ \bibinfo {author} {\bibfnamefont
  {D.}~\bibnamefont {Gershoni}},\ }\bibfield  {title} {\bibinfo {title}
  {Deterministic generation of a cluster state of entangled photons},\ }\href
  {https://doi.org/10.1126/science.aah4758} {\bibfield  {journal} {\bibinfo
  {journal} {Science}\ }\textbf {\bibinfo {volume} {354}},\ \bibinfo {pages}
  {434} (\bibinfo {year} {2016})}\BibitemShut {NoStop}%
\bibitem [{\citenamefont {Gangloff}\ \emph {et~al.}(2019)\citenamefont
  {Gangloff}, \citenamefont {Éthier Majcher}, \citenamefont {Lang},
  \citenamefont {Denning}, \citenamefont {Bodey}, \citenamefont {Jackson},
  \citenamefont {Clarke}, \citenamefont {Hugues}, \citenamefont {Gall},\ and\
  \citenamefont {Atatüre}}]{gangloff2019}%
  \BibitemOpen
  \bibfield  {author} {\bibinfo {author} {\bibfnamefont {D.~A.}\ \bibnamefont
  {Gangloff}}, \bibinfo {author} {\bibfnamefont {G.}~\bibnamefont {Éthier
  Majcher}}, \bibinfo {author} {\bibfnamefont {C.}~\bibnamefont {Lang}},
  \bibinfo {author} {\bibfnamefont {E.~V.}\ \bibnamefont {Denning}}, \bibinfo
  {author} {\bibfnamefont {J.~H.}\ \bibnamefont {Bodey}}, \bibinfo {author}
  {\bibfnamefont {D.~M.}\ \bibnamefont {Jackson}}, \bibinfo {author}
  {\bibfnamefont {E.}~\bibnamefont {Clarke}}, \bibinfo {author} {\bibfnamefont
  {M.}~\bibnamefont {Hugues}}, \bibinfo {author} {\bibfnamefont {C.~L.}\
  \bibnamefont {Gall}},\ and\ \bibinfo {author} {\bibfnamefont
  {M.}~\bibnamefont {Atatüre}},\ }\bibfield  {title} {\bibinfo {title}
  {Quantum interface of an electron and a nuclear ensemble},\ }\href
  {https://doi.org/10.1126/science.aaw2906} {\bibfield  {journal} {\bibinfo
  {journal} {Science}\ }\textbf {\bibinfo {volume} {364}},\ \bibinfo {pages}
  {62} (\bibinfo {year} {2019})}\BibitemShut {NoStop}%
\bibitem [{\citenamefont {Cabot}\ \emph {et~al.}(2022)\citenamefont {Cabot},
  \citenamefont {Carollo},\ and\ \citenamefont {Lesanovsky}}]{cabot2022}%
  \BibitemOpen
  \bibfield  {author} {\bibinfo {author} {\bibfnamefont {A.}~\bibnamefont
  {Cabot}}, \bibinfo {author} {\bibfnamefont {F.}~\bibnamefont {Carollo}},\
  and\ \bibinfo {author} {\bibfnamefont {I.}~\bibnamefont {Lesanovsky}},\
  }\bibfield  {title} {\bibinfo {title} {Metastable discrete time-crystal
  resonances in a dissipative central spin system},\ }\href
  {https://doi.org/10.1103/PhysRevB.106.134311} {\bibfield  {journal} {\bibinfo
   {journal} {Phys. Rev. B}\ }\textbf {\bibinfo {volume} {106}},\ \bibinfo
  {pages} {134311} (\bibinfo {year} {2022})}\BibitemShut {NoStop}%
\bibitem [{\citenamefont {Greilich}\ \emph {et~al.}(2023)\citenamefont
  {Greilich}, \citenamefont {Kopteva}, \citenamefont {Kamenskii}, \citenamefont
  {Sokolov}, \citenamefont {Korenev},\ and\ \citenamefont
  {Bayer}}]{greilich2023}%
  \BibitemOpen
  \bibfield  {author} {\bibinfo {author} {\bibfnamefont {A.}~\bibnamefont
  {Greilich}}, \bibinfo {author} {\bibfnamefont {N.~E.}\ \bibnamefont
  {Kopteva}}, \bibinfo {author} {\bibfnamefont {A.~N.}\ \bibnamefont
  {Kamenskii}}, \bibinfo {author} {\bibfnamefont {P.~S.}\ \bibnamefont
  {Sokolov}}, \bibinfo {author} {\bibfnamefont {V.~L.}\ \bibnamefont
  {Korenev}},\ and\ \bibinfo {author} {\bibfnamefont {M.}~\bibnamefont
  {Bayer}},\ }\bibfield  {title} {\bibinfo {title} {Continuous time crystal in
  an electron-nuclear spin system: stability and melting of periodic
  auto-oscillations},\ }\href {https://doi.org/10.48550/arXiv.2303.15989}
  {\bibfield  {journal} {\bibinfo  {journal} {arXiv:2303.15989}\ } (\bibinfo
  {year} {2023})}\BibitemShut {NoStop}%
\bibitem [{\citenamefont {Schliemann}\ \emph {et~al.}(2003)\citenamefont
  {Schliemann}, \citenamefont {Khaetskii},\ and\ \citenamefont
  {Loss}}]{schliemann2003}%
  \BibitemOpen
  \bibfield  {author} {\bibinfo {author} {\bibfnamefont {J.}~\bibnamefont
  {Schliemann}}, \bibinfo {author} {\bibfnamefont {A.}~\bibnamefont
  {Khaetskii}},\ and\ \bibinfo {author} {\bibfnamefont {D.}~\bibnamefont
  {Loss}},\ }\bibfield  {title} {\bibinfo {title} {Electron spin dynamics in
  quantum dots and related nanostructures due to hyperfine interaction with
  nuclei},\ }\href {https://doi.org/10.1088/0953-8984/15/50/R01} {\bibfield
  {journal} {\bibinfo  {journal} {J. Phys.: Condens. Matter}\ }\textbf
  {\bibinfo {volume} {15}},\ \bibinfo {pages} {R1809} (\bibinfo {year}
  {2003})}\BibitemShut {NoStop}%
\bibitem [{\citenamefont {Taylor}\ \emph {et~al.}(2003)\citenamefont {Taylor},
  \citenamefont {Marcus},\ and\ \citenamefont {Lukin}}]{taylor2003}%
  \BibitemOpen
  \bibfield  {author} {\bibinfo {author} {\bibfnamefont {J.~M.}\ \bibnamefont
  {Taylor}}, \bibinfo {author} {\bibfnamefont {C.~M.}\ \bibnamefont {Marcus}},\
  and\ \bibinfo {author} {\bibfnamefont {M.~D.}\ \bibnamefont {Lukin}},\
  }\bibfield  {title} {\bibinfo {title} {{Long-Lived Memory for Mesoscopic
  Quantum Bits}},\ }\href {https://doi.org/10.1103/PhysRevLett.90.206803}
  {\bibfield  {journal} {\bibinfo  {journal} {Phys. Rev. Lett.}\ }\textbf
  {\bibinfo {volume} {90}},\ \bibinfo {pages} {206803} (\bibinfo {year}
  {2003})}\BibitemShut {NoStop}%
\bibitem [{\citenamefont {Togan}\ \emph {et~al.}(2011)\citenamefont {Togan},
  \citenamefont {Chu}, \citenamefont {Imamoglu},\ and\ \citenamefont
  {Lukin}}]{togan2011}%
  \BibitemOpen
  \bibfield  {author} {\bibinfo {author} {\bibfnamefont {E.}~\bibnamefont
  {Togan}}, \bibinfo {author} {\bibfnamefont {Y.}~\bibnamefont {Chu}}, \bibinfo
  {author} {\bibfnamefont {A.}~\bibnamefont {Imamoglu}},\ and\ \bibinfo
  {author} {\bibfnamefont {M.~D.}\ \bibnamefont {Lukin}},\ }\bibfield  {title}
  {\bibinfo {title} {Laser cooling and real-time measurement of the nuclear
  spin environment of a solid-state qubit},\ }\href
  {https://doi.org/10.1038/nature10528} {\bibfield  {journal} {\bibinfo
  {journal} {Nature}\ }\textbf {\bibinfo {volume} {478}},\ \bibinfo {pages}
  {497} (\bibinfo {year} {2011})}\BibitemShut {NoStop}%
\bibitem [{\citenamefont {Urbaszek}\ \emph {et~al.}(2013)\citenamefont
  {Urbaszek}, \citenamefont {Marie}, \citenamefont {Amand}, \citenamefont
  {Krebs}, \citenamefont {Voisin}, \citenamefont {Maletinsky}, \citenamefont
  {H\"ogele},\ and\ \citenamefont {Imamoglu}}]{urbaszek2013}%
  \BibitemOpen
  \bibfield  {author} {\bibinfo {author} {\bibfnamefont {B.}~\bibnamefont
  {Urbaszek}}, \bibinfo {author} {\bibfnamefont {X.}~\bibnamefont {Marie}},
  \bibinfo {author} {\bibfnamefont {T.}~\bibnamefont {Amand}}, \bibinfo
  {author} {\bibfnamefont {O.}~\bibnamefont {Krebs}}, \bibinfo {author}
  {\bibfnamefont {P.}~\bibnamefont {Voisin}}, \bibinfo {author} {\bibfnamefont
  {P.}~\bibnamefont {Maletinsky}}, \bibinfo {author} {\bibfnamefont
  {A.}~\bibnamefont {H\"ogele}},\ and\ \bibinfo {author} {\bibfnamefont
  {A.}~\bibnamefont {Imamoglu}},\ }\bibfield  {title} {\bibinfo {title}
  {{Nuclear spin physics in quantum dots: An optical investigation}},\ }\href
  {https://doi.org/10.1103/RevModPhys.85.79} {\bibfield  {journal} {\bibinfo
  {journal} {Rev. Mod. Phys.}\ }\textbf {\bibinfo {volume} {85}},\ \bibinfo
  {pages} {79} (\bibinfo {year} {2013})}\BibitemShut {NoStop}%
\bibitem [{\citenamefont {{Lilly Thankamony}}\ \emph
  {et~al.}(2017)\citenamefont {{Lilly Thankamony}}, \citenamefont {Wittmann},
  \citenamefont {Kaushik},\ and\ \citenamefont {Corzilius}}]{aany2017}%
  \BibitemOpen
  \bibfield  {author} {\bibinfo {author} {\bibfnamefont {A.~S.}\ \bibnamefont
  {{Lilly Thankamony}}}, \bibinfo {author} {\bibfnamefont {J.~J.}\ \bibnamefont
  {Wittmann}}, \bibinfo {author} {\bibfnamefont {M.}~\bibnamefont {Kaushik}},\
  and\ \bibinfo {author} {\bibfnamefont {B.}~\bibnamefont {Corzilius}},\
  }\bibfield  {title} {\bibinfo {title} {{Dynamic nuclear polarization for
  sensitivity enhancement in modern solid-state NMR}},\ }\href
  {https://doi.org/https://doi.org/10.1016/j.pnmrs.2017.06.002} {\bibfield
  {journal} {\bibinfo  {journal} {Prog. Nucl. Magn. Reson. Spectrosc.}\
  }\textbf {\bibinfo {volume} {102-103}},\ \bibinfo {pages} {120} (\bibinfo
  {year} {2017})}\BibitemShut {NoStop}%
\bibitem [{\citenamefont {Fernández-Acebal}\ \emph {et~al.}(2018)\citenamefont
  {Fernández-Acebal}, \citenamefont {Rosolio}, \citenamefont {Scheuer},
  \citenamefont {Müller}, \citenamefont {Müller}, \citenamefont {Schmitt},
  \citenamefont {McGuinness}, \citenamefont {Schwarz}, \citenamefont {Chen},
  \citenamefont {Retzker}, \citenamefont {Naydenov}, \citenamefont {Jelezko},\
  and\ \citenamefont {Plenio}}]{fernandez2018}%
  \BibitemOpen
  \bibfield  {author} {\bibinfo {author} {\bibfnamefont {P.}~\bibnamefont
  {Fernández-Acebal}}, \bibinfo {author} {\bibfnamefont {O.}~\bibnamefont
  {Rosolio}}, \bibinfo {author} {\bibfnamefont {J.}~\bibnamefont {Scheuer}},
  \bibinfo {author} {\bibfnamefont {C.}~\bibnamefont {Müller}}, \bibinfo
  {author} {\bibfnamefont {S.}~\bibnamefont {Müller}}, \bibinfo {author}
  {\bibfnamefont {S.}~\bibnamefont {Schmitt}}, \bibinfo {author} {\bibfnamefont
  {L.}~\bibnamefont {McGuinness}}, \bibinfo {author} {\bibfnamefont
  {I.}~\bibnamefont {Schwarz}}, \bibinfo {author} {\bibfnamefont
  {Q.}~\bibnamefont {Chen}}, \bibinfo {author} {\bibfnamefont {A.}~\bibnamefont
  {Retzker}}, \bibinfo {author} {\bibfnamefont {B.}~\bibnamefont {Naydenov}},
  \bibinfo {author} {\bibfnamefont {F.}~\bibnamefont {Jelezko}},\ and\ \bibinfo
  {author} {\bibfnamefont {M.}~\bibnamefont {Plenio}},\ }\bibfield  {title}
  {\bibinfo {title} {{Toward Hyperpolarization of Oil Molecules via Single
  Nitrogen Vacancy Centers in Diamond}},\ }\href
  {https://doi.org/10.1021/acs.nanolett.7b05175} {\bibfield  {journal}
  {\bibinfo  {journal} {Nano Lett.}\ }\textbf {\bibinfo {volume} {18}},\
  \bibinfo {pages} {1882} (\bibinfo {year} {2018})}\BibitemShut {NoStop}%
\bibitem [{\citenamefont {Villazon}\ \emph {et~al.}(2021)\citenamefont
  {Villazon}, \citenamefont {Claeys}, \citenamefont {Polkovnikov},\ and\
  \citenamefont {Chandran}}]{villazon2021}%
  \BibitemOpen
  \bibfield  {author} {\bibinfo {author} {\bibfnamefont {T.}~\bibnamefont
  {Villazon}}, \bibinfo {author} {\bibfnamefont {P.~W.}\ \bibnamefont
  {Claeys}}, \bibinfo {author} {\bibfnamefont {A.}~\bibnamefont
  {Polkovnikov}},\ and\ \bibinfo {author} {\bibfnamefont {A.}~\bibnamefont
  {Chandran}},\ }\bibfield  {title} {\bibinfo {title} {Shortcuts to dynamic
  polarization},\ }\href {https://doi.org/10.1103/PhysRevB.103.075118}
  {\bibfield  {journal} {\bibinfo  {journal} {Phys. Rev. B}\ }\textbf {\bibinfo
  {volume} {103}},\ \bibinfo {pages} {075118} (\bibinfo {year}
  {2021})}\BibitemShut {NoStop}%
\bibitem [{\citenamefont {Rizzato}\ \emph {et~al.}(2022)\citenamefont
  {Rizzato}, \citenamefont {Bruckmaier}, \citenamefont {Liu}, \citenamefont
  {Glaser},\ and\ \citenamefont {Bucher}}]{rizzato2022}%
  \BibitemOpen
  \bibfield  {author} {\bibinfo {author} {\bibfnamefont {R.}~\bibnamefont
  {Rizzato}}, \bibinfo {author} {\bibfnamefont {F.}~\bibnamefont {Bruckmaier}},
  \bibinfo {author} {\bibfnamefont {K.}~\bibnamefont {Liu}}, \bibinfo {author}
  {\bibfnamefont {S.}~\bibnamefont {Glaser}},\ and\ \bibinfo {author}
  {\bibfnamefont {D.}~\bibnamefont {Bucher}},\ }\bibfield  {title} {\bibinfo
  {title} {{Polarization Transfer from Optically Pumped Ensembles of N-$V$
  Centers to Multinuclear Spin Baths}},\ }\href
  {https://doi.org/10.1103/PhysRevApplied.17.024067} {\bibfield  {journal}
  {\bibinfo  {journal} {Phys. Rev. Appl.}\ }\textbf {\bibinfo {volume} {17}},\
  \bibinfo {pages} {024067} (\bibinfo {year} {2022})}\BibitemShut {NoStop}%
\bibitem [{\citenamefont {Allert}\ \emph {et~al.}(2022)\citenamefont {Allert},
  \citenamefont {Briegel},\ and\ \citenamefont {Bucher}}]{allert2022}%
  \BibitemOpen
  \bibfield  {author} {\bibinfo {author} {\bibfnamefont {R.~D.}\ \bibnamefont
  {Allert}}, \bibinfo {author} {\bibfnamefont {K.~D.}\ \bibnamefont
  {Briegel}},\ and\ \bibinfo {author} {\bibfnamefont {D.~B.}\ \bibnamefont
  {Bucher}},\ }\bibfield  {title} {\bibinfo {title} {{Advances in nano- and
  microscale NMR spectroscopy using diamond quantum sensors}},\ }\href
  {https://doi.org/10.1039/D2CC01546C} {\bibfield  {journal} {\bibinfo
  {journal} {Chem. Commun.}\ }\textbf {\bibinfo {volume} {58}},\ \bibinfo
  {pages} {8165} (\bibinfo {year} {2022})}\BibitemShut {NoStop}%
\bibitem [{\citenamefont {Yang}\ and\ \citenamefont {Liu}(2008)}]{yang2008}%
  \BibitemOpen
  \bibfield  {author} {\bibinfo {author} {\bibfnamefont {W.}~\bibnamefont
  {Yang}}\ and\ \bibinfo {author} {\bibfnamefont {R.-B.}\ \bibnamefont {Liu}},\
  }\bibfield  {title} {\bibinfo {title} {Quantum many-body theory of qubit
  decoherence in a finite-size spin bath},\ }\href
  {https://doi.org/10.1103/PhysRevB.78.085315} {\bibfield  {journal} {\bibinfo
  {journal} {Phys. Rev. B}\ }\textbf {\bibinfo {volume} {78}},\ \bibinfo
  {pages} {085315} (\bibinfo {year} {2008})}\BibitemShut {NoStop}%
\bibitem [{\citenamefont {Chekhovich}\ \emph {et~al.}(2013)\citenamefont
  {Chekhovich}, \citenamefont {Makhonin}, \citenamefont {Tartakovskii},
  \citenamefont {Yacoby}, \citenamefont {Bluhm}, \citenamefont {Nowack},\ and\
  \citenamefont {Vandersypen}}]{chekhovich2013}%
  \BibitemOpen
  \bibfield  {author} {\bibinfo {author} {\bibfnamefont {E.~A.}\ \bibnamefont
  {Chekhovich}}, \bibinfo {author} {\bibfnamefont {M.~N.}\ \bibnamefont
  {Makhonin}}, \bibinfo {author} {\bibfnamefont {A.~I.}\ \bibnamefont
  {Tartakovskii}}, \bibinfo {author} {\bibfnamefont {A.}~\bibnamefont
  {Yacoby}}, \bibinfo {author} {\bibfnamefont {H.}~\bibnamefont {Bluhm}},
  \bibinfo {author} {\bibfnamefont {K.~C.}\ \bibnamefont {Nowack}},\ and\
  \bibinfo {author} {\bibfnamefont {L.~M.~K.}\ \bibnamefont {Vandersypen}},\
  }\bibfield  {title} {\bibinfo {title} {Nuclear spin effects in semiconductor
  quantum dots},\ }\href {https://doi.org/10.1038/nmat3652} {\bibfield
  {journal} {\bibinfo  {journal} {Nat. Mater.}\ }\textbf {\bibinfo {volume}
  {12}},\ \bibinfo {pages} {494} (\bibinfo {year} {2013})}\BibitemShut
  {NoStop}%
\bibitem [{\citenamefont {Lindoy}\ and\ \citenamefont
  {Manolopoulos}(2018)}]{lindoy2018}%
  \BibitemOpen
  \bibfield  {author} {\bibinfo {author} {\bibfnamefont {L.~P.}\ \bibnamefont
  {Lindoy}}\ and\ \bibinfo {author} {\bibfnamefont {D.~E.}\ \bibnamefont
  {Manolopoulos}},\ }\bibfield  {title} {\bibinfo {title} {{Simple and Accurate
  Method for Central Spin Problems}},\ }\href
  {https://doi.org/10.1103/PhysRevLett.120.220604} {\bibfield  {journal}
  {\bibinfo  {journal} {Phys. Rev. Lett.}\ }\textbf {\bibinfo {volume} {120}},\
  \bibinfo {pages} {220604} (\bibinfo {year} {2018})}\BibitemShut {NoStop}%
\bibitem [{\citenamefont {R\"ohrig}\ \emph {et~al.}(2018)\citenamefont
  {R\"ohrig}, \citenamefont {Schering}, \citenamefont {Gravert}, \citenamefont
  {Fauseweh},\ and\ \citenamefont {Uhrig}}]{rohrig2018}%
  \BibitemOpen
  \bibfield  {author} {\bibinfo {author} {\bibfnamefont {R.}~\bibnamefont
  {R\"ohrig}}, \bibinfo {author} {\bibfnamefont {P.}~\bibnamefont {Schering}},
  \bibinfo {author} {\bibfnamefont {L.~B.}\ \bibnamefont {Gravert}}, \bibinfo
  {author} {\bibfnamefont {B.}~\bibnamefont {Fauseweh}},\ and\ \bibinfo
  {author} {\bibfnamefont {G.~S.}\ \bibnamefont {Uhrig}},\ }\bibfield  {title}
  {\bibinfo {title} {Quantum mechanical treatment of large spin baths},\ }\href
  {https://doi.org/10.1103/PhysRevB.97.165431} {\bibfield  {journal} {\bibinfo
  {journal} {Phys. Rev. B}\ }\textbf {\bibinfo {volume} {97}},\ \bibinfo
  {pages} {165431} (\bibinfo {year} {2018})}\BibitemShut {NoStop}%
\bibitem [{\citenamefont {Fowler-Wright}\ \emph {et~al.}(2023)\citenamefont
  {Fowler-Wright}, \citenamefont {Arnardóttir}, \citenamefont {Kirton},
  \citenamefont {Lovett},\ and\ \citenamefont {Keeling}}]{fowler2023}%
  \BibitemOpen
  \bibfield  {author} {\bibinfo {author} {\bibfnamefont {P.}~\bibnamefont
  {Fowler-Wright}}, \bibinfo {author} {\bibfnamefont {K.~B.}\ \bibnamefont
  {Arnardóttir}}, \bibinfo {author} {\bibfnamefont {P.}~\bibnamefont
  {Kirton}}, \bibinfo {author} {\bibfnamefont {B.~W.}\ \bibnamefont {Lovett}},\
  and\ \bibinfo {author} {\bibfnamefont {J.}~\bibnamefont {Keeling}},\
  }\bibfield  {title} {\bibinfo {title} {Determining the validity of cumulant
  expansions for central spin models},\ }\href
  {https://doi.org/10.48550/arXiv.2303.04410} {\bibfield  {journal} {\bibinfo
  {journal} {arXiv:2303.04410}\ } (\bibinfo {year} {2023})}\BibitemShut
  {NoStop}%
\bibitem [{\citenamefont {Rudner}\ \emph {et~al.}(2011)\citenamefont {Rudner},
  \citenamefont {Vandersypen}, \citenamefont {Vuleti\ifmmode~\acute{c}\else
  \'{c}\fi{}},\ and\ \citenamefont {Levitov}}]{rudner2011}%
  \BibitemOpen
  \bibfield  {author} {\bibinfo {author} {\bibfnamefont {M.~S.}\ \bibnamefont
  {Rudner}}, \bibinfo {author} {\bibfnamefont {L.~M.~K.}\ \bibnamefont
  {Vandersypen}}, \bibinfo {author} {\bibfnamefont {V.}~\bibnamefont
  {Vuleti\ifmmode~\acute{c}\else \'{c}\fi{}}},\ and\ \bibinfo {author}
  {\bibfnamefont {L.~S.}\ \bibnamefont {Levitov}},\ }\bibfield  {title}
  {\bibinfo {title} {{Generating Entanglement and Squeezed States of Nuclear
  Spins in Quantum Dots}},\ }\href
  {https://doi.org/10.1103/PhysRevLett.107.206806} {\bibfield  {journal}
  {\bibinfo  {journal} {Phys. Rev. Lett.}\ }\textbf {\bibinfo {volume} {107}},\
  \bibinfo {pages} {206806} (\bibinfo {year} {2011})}\BibitemShut {NoStop}%
\bibitem [{\citenamefont {de~Lange}\ \emph {et~al.}(2012)\citenamefont
  {de~Lange}, \citenamefont {van~der Sar}, \citenamefont {Blok}, \citenamefont
  {Wang}, \citenamefont {Dobrovitski},\ and\ \citenamefont
  {Hanson}}]{delange2012}%
  \BibitemOpen
  \bibfield  {author} {\bibinfo {author} {\bibfnamefont {G.}~\bibnamefont
  {de~Lange}}, \bibinfo {author} {\bibfnamefont {T.}~\bibnamefont {van~der
  Sar}}, \bibinfo {author} {\bibfnamefont {M.}~\bibnamefont {Blok}}, \bibinfo
  {author} {\bibfnamefont {Z.-H.}\ \bibnamefont {Wang}}, \bibinfo {author}
  {\bibfnamefont {V.}~\bibnamefont {Dobrovitski}},\ and\ \bibinfo {author}
  {\bibfnamefont {R.}~\bibnamefont {Hanson}},\ }\bibfield  {title} {\bibinfo
  {title} {Controlling the quantum dynamics of a mesoscopic spin bath in
  diamond},\ }\href {https://doi.org/10.1038/srep00382} {\bibfield  {journal}
  {\bibinfo  {journal} {Sci. Rep.}\ }\textbf {\bibinfo {volume} {2}},\ \bibinfo
  {pages} {382} (\bibinfo {year} {2012})}\BibitemShut {NoStop}%
\bibitem [{\citenamefont {Gao}\ \emph {et~al.}(2015)\citenamefont {Gao},
  \citenamefont {Imamoglu}, \citenamefont {Bernien},\ and\ \citenamefont
  {Hanson}}]{gao2015}%
  \BibitemOpen
  \bibfield  {author} {\bibinfo {author} {\bibfnamefont {W.~B.}\ \bibnamefont
  {Gao}}, \bibinfo {author} {\bibfnamefont {A.}~\bibnamefont {Imamoglu}},
  \bibinfo {author} {\bibfnamefont {H.}~\bibnamefont {Bernien}},\ and\ \bibinfo
  {author} {\bibfnamefont {R.}~\bibnamefont {Hanson}},\ }\bibfield  {title}
  {\bibinfo {title} {Coherent manipulation, measurement and entanglement of
  individual solid-state spins using optical fields},\ }\href
  {https://doi.org/10.1038/nphoton.2015.58} {\bibfield  {journal} {\bibinfo
  {journal} {Nat. Photonics}\ }\textbf {\bibinfo {volume} {9}},\ \bibinfo
  {pages} {363} (\bibinfo {year} {2015})}\BibitemShut {NoStop}%
\bibitem [{\citenamefont {Bauch}\ \emph {et~al.}(2018)\citenamefont {Bauch},
  \citenamefont {Hart}, \citenamefont {Schloss}, \citenamefont {Turner},
  \citenamefont {Barry}, \citenamefont {Kehayias}, \citenamefont {Singh},\ and\
  \citenamefont {Walsworth}}]{bauch2018}%
  \BibitemOpen
  \bibfield  {author} {\bibinfo {author} {\bibfnamefont {E.}~\bibnamefont
  {Bauch}}, \bibinfo {author} {\bibfnamefont {C.~A.}\ \bibnamefont {Hart}},
  \bibinfo {author} {\bibfnamefont {J.~M.}\ \bibnamefont {Schloss}}, \bibinfo
  {author} {\bibfnamefont {M.~J.}\ \bibnamefont {Turner}}, \bibinfo {author}
  {\bibfnamefont {J.~F.}\ \bibnamefont {Barry}}, \bibinfo {author}
  {\bibfnamefont {P.}~\bibnamefont {Kehayias}}, \bibinfo {author}
  {\bibfnamefont {S.}~\bibnamefont {Singh}},\ and\ \bibinfo {author}
  {\bibfnamefont {R.~L.}\ \bibnamefont {Walsworth}},\ }\bibfield  {title}
  {\bibinfo {title} {{Ultralong Dephasing Times in Solid-State Spin Ensembles
  via Quantum Control}},\ }\href {https://doi.org/10.1103/PhysRevX.8.031025}
  {\bibfield  {journal} {\bibinfo  {journal} {Phys. Rev. X}\ }\textbf {\bibinfo
  {volume} {8}},\ \bibinfo {pages} {031025} (\bibinfo {year}
  {2018})}\BibitemShut {NoStop}%
\bibitem [{\citenamefont {Chen}\ \emph {et~al.}(2018)\citenamefont {Chen},
  \citenamefont {Zopf}, \citenamefont {Keil}, \citenamefont {Ding},\ and\
  \citenamefont {Schmidt}}]{chen2018}%
  \BibitemOpen
  \bibfield  {author} {\bibinfo {author} {\bibfnamefont {Y.}~\bibnamefont
  {Chen}}, \bibinfo {author} {\bibfnamefont {M.}~\bibnamefont {Zopf}}, \bibinfo
  {author} {\bibfnamefont {R.}~\bibnamefont {Keil}}, \bibinfo {author}
  {\bibfnamefont {F.}~\bibnamefont {Ding}},\ and\ \bibinfo {author}
  {\bibfnamefont {O.~G.}\ \bibnamefont {Schmidt}},\ }\bibfield  {title}
  {\bibinfo {title} {Highly-efficient extraction of entangled photons from
  quantum dots using a broadband optical antenna},\ }\href
  {https://doi.org/10.1038/s41467-018-05456-2} {\bibfield  {journal} {\bibinfo
  {journal} {Nat. Commun.}\ }\textbf {\bibinfo {volume} {9}},\ \bibinfo {pages}
  {2994} (\bibinfo {year} {2018})}\BibitemShut {NoStop}%
\bibitem [{\citenamefont {Unden}\ \emph {et~al.}(2018)\citenamefont {Unden},
  \citenamefont {Tomek}, \citenamefont {Weggler}, \citenamefont {Frank},
  \citenamefont {London}, \citenamefont {Zopes}, \citenamefont {Degen},
  \citenamefont {Raatz}, \citenamefont {Meijer}, \citenamefont {Watanabe},
  \citenamefont {Itoh}, \citenamefont {Plenio}, \citenamefont {Naydenov},\ and\
  \citenamefont {Jelezko}}]{unden2018}%
  \BibitemOpen
  \bibfield  {author} {\bibinfo {author} {\bibfnamefont {T.}~\bibnamefont
  {Unden}}, \bibinfo {author} {\bibfnamefont {N.}~\bibnamefont {Tomek}},
  \bibinfo {author} {\bibfnamefont {T.}~\bibnamefont {Weggler}}, \bibinfo
  {author} {\bibfnamefont {F.}~\bibnamefont {Frank}}, \bibinfo {author}
  {\bibfnamefont {P.}~\bibnamefont {London}}, \bibinfo {author} {\bibfnamefont
  {J.}~\bibnamefont {Zopes}}, \bibinfo {author} {\bibfnamefont
  {C.}~\bibnamefont {Degen}}, \bibinfo {author} {\bibfnamefont
  {N.}~\bibnamefont {Raatz}}, \bibinfo {author} {\bibfnamefont
  {J.}~\bibnamefont {Meijer}}, \bibinfo {author} {\bibfnamefont
  {H.}~\bibnamefont {Watanabe}}, \bibinfo {author} {\bibfnamefont {K.~M.}\
  \bibnamefont {Itoh}}, \bibinfo {author} {\bibfnamefont {M.~B.}\ \bibnamefont
  {Plenio}}, \bibinfo {author} {\bibfnamefont {B.}~\bibnamefont {Naydenov}},\
  and\ \bibinfo {author} {\bibfnamefont {F.}~\bibnamefont {Jelezko}},\
  }\bibfield  {title} {\bibinfo {title} {Coherent control of solid state
  nuclear spin nano-ensembles},\ }\href
  {https://doi.org/10.1038/s41534-018-0089-8} {\bibfield  {journal} {\bibinfo
  {journal} {npj Quantum Inf.}\ }\textbf {\bibinfo {volume} {4}},\ \bibinfo
  {pages} {39} (\bibinfo {year} {2018})}\BibitemShut {NoStop}%
\bibitem [{\citenamefont {Dong}\ \emph {et~al.}(2019)\citenamefont {Dong},
  \citenamefont {Liang}, \citenamefont {Duan}, \citenamefont {Wang},
  \citenamefont {Li}, \citenamefont {Rong},\ and\ \citenamefont
  {Du}}]{dong2019}%
  \BibitemOpen
  \bibfield  {author} {\bibinfo {author} {\bibfnamefont {L.}~\bibnamefont
  {Dong}}, \bibinfo {author} {\bibfnamefont {H.}~\bibnamefont {Liang}},
  \bibinfo {author} {\bibfnamefont {C.-K.}\ \bibnamefont {Duan}}, \bibinfo
  {author} {\bibfnamefont {Y.}~\bibnamefont {Wang}}, \bibinfo {author}
  {\bibfnamefont {Z.}~\bibnamefont {Li}}, \bibinfo {author} {\bibfnamefont
  {X.}~\bibnamefont {Rong}},\ and\ \bibinfo {author} {\bibfnamefont
  {J.}~\bibnamefont {Du}},\ }\bibfield  {title} {\bibinfo {title} {Optimal
  control of a spin bath},\ }\href {https://doi.org/10.1103/PhysRevA.99.013426}
  {\bibfield  {journal} {\bibinfo  {journal} {Phys. Rev. A}\ }\textbf {\bibinfo
  {volume} {99}},\ \bibinfo {pages} {013426} (\bibinfo {year}
  {2019})}\BibitemShut {NoStop}%
\bibitem [{\citenamefont {Liu}\ \emph {et~al.}(2021)\citenamefont {Liu},
  \citenamefont {Shi}, \citenamefont {Shi}, \citenamefont {Wang},\ and\
  \citenamefont {Yang}}]{liu2021}%
  \BibitemOpen
  \bibfield  {author} {\bibinfo {author} {\bibfnamefont {J.-X.}\ \bibnamefont
  {Liu}}, \bibinfo {author} {\bibfnamefont {H.-L.}\ \bibnamefont {Shi}},
  \bibinfo {author} {\bibfnamefont {Y.-H.}\ \bibnamefont {Shi}}, \bibinfo
  {author} {\bibfnamefont {X.-H.}\ \bibnamefont {Wang}},\ and\ \bibinfo
  {author} {\bibfnamefont {W.-L.}\ \bibnamefont {Yang}},\ }\bibfield  {title}
  {\bibinfo {title} {Entanglement and work extraction in the central-spin
  quantum battery},\ }\href {https://doi.org/10.1103/PhysRevB.104.245418}
  {\bibfield  {journal} {\bibinfo  {journal} {Phys. Rev. B}\ }\textbf {\bibinfo
  {volume} {104}},\ \bibinfo {pages} {245418} (\bibinfo {year}
  {2021})}\BibitemShut {NoStop}%
\bibitem [{\citenamefont {Degen}\ \emph {et~al.}(2021)\citenamefont {Degen},
  \citenamefont {Loenen}, \citenamefont {Bartling}, \citenamefont {Bradley},
  \citenamefont {Meinsma}, \citenamefont {Markham}, \citenamefont {Twitchen},\
  and\ \citenamefont {Taminiau}}]{degen2021}%
  \BibitemOpen
  \bibfield  {author} {\bibinfo {author} {\bibfnamefont {M.~J.}\ \bibnamefont
  {Degen}}, \bibinfo {author} {\bibfnamefont {S.~J.~H.}\ \bibnamefont
  {Loenen}}, \bibinfo {author} {\bibfnamefont {H.~P.}\ \bibnamefont
  {Bartling}}, \bibinfo {author} {\bibfnamefont {C.~E.}\ \bibnamefont
  {Bradley}}, \bibinfo {author} {\bibfnamefont {A.~L.}\ \bibnamefont
  {Meinsma}}, \bibinfo {author} {\bibfnamefont {M.}~\bibnamefont {Markham}},
  \bibinfo {author} {\bibfnamefont {D.~J.}\ \bibnamefont {Twitchen}},\ and\
  \bibinfo {author} {\bibfnamefont {T.~H.}\ \bibnamefont {Taminiau}},\
  }\bibfield  {title} {\bibinfo {title} {Entanglement of dark electron-nuclear
  spin defects in diamond},\ }\href
  {https://doi.org/10.1038/s41467-021-23454-9} {\bibfield  {journal} {\bibinfo
  {journal} {Nat. Commun.}\ }\textbf {\bibinfo {volume} {12}},\ \bibinfo
  {pages} {3470} (\bibinfo {year} {2021})}\BibitemShut {NoStop}%
\bibitem [{\citenamefont {Gillard}\ \emph {et~al.}(2022)\citenamefont
  {Gillard}, \citenamefont {Clarke},\ and\ \citenamefont
  {Chekhovich}}]{gillard2022}%
  \BibitemOpen
  \bibfield  {author} {\bibinfo {author} {\bibfnamefont {G.}~\bibnamefont
  {Gillard}}, \bibinfo {author} {\bibfnamefont {E.}~\bibnamefont {Clarke}},\
  and\ \bibinfo {author} {\bibfnamefont {E.~A.}\ \bibnamefont {Chekhovich}},\
  }\bibfield  {title} {\bibinfo {title} {Harnessing many-body spin environment
  for long coherence storage and high-fidelity single-shot qubit readout},\
  }\href {https://doi.org/10.1038/s41467-022-31618-4} {\bibfield  {journal}
  {\bibinfo  {journal} {Nat. Commun.}\ }\textbf {\bibinfo {volume} {13}},\
  \bibinfo {pages} {4048} (\bibinfo {year} {2022})}\BibitemShut {NoStop}%
\bibitem [{\citenamefont {Jaynes}\ and\ \citenamefont
  {Cummings}(1963)}]{jaynes1963}%
  \BibitemOpen
  \bibfield  {author} {\bibinfo {author} {\bibfnamefont {E.}~\bibnamefont
  {Jaynes}}\ and\ \bibinfo {author} {\bibfnamefont {F.}~\bibnamefont
  {Cummings}},\ }\bibfield  {title} {\bibinfo {title} {{Comparison of quantum
  and semiclassical radiation theories with application to the beam maser}},\
  }\href {https://doi.org/10.1109/PROC.1963.1664} {\bibfield  {journal}
  {\bibinfo  {journal} {Proc. IEEE}\ }\textbf {\bibinfo {volume} {51}},\
  \bibinfo {pages} {89} (\bibinfo {year} {1963})}\BibitemShut {NoStop}%
\bibitem [{\citenamefont {Goderis}\ and\ \citenamefont
  {Vets}(1989)}]{goderis1989}%
  \BibitemOpen
  \bibfield  {author} {\bibinfo {author} {\bibfnamefont {D.}~\bibnamefont
  {Goderis}}\ and\ \bibinfo {author} {\bibfnamefont {P.}~\bibnamefont {Vets}},\
  }\bibfield  {title} {\bibinfo {title} {{Central limit theorem for mixing
  quantum systems and the CCR-algebra of fluctuations}},\ }\href
  {https://doi.org/10.1007/BF01257415} {\bibfield  {journal} {\bibinfo
  {journal} {Commun. Math. Phys.}\ }\textbf {\bibinfo {volume} {122}},\
  \bibinfo {pages} {249} (\bibinfo {year} {1989})}\BibitemShut {NoStop}%
\bibitem [{\citenamefont {Goderis}\ \emph {et~al.}(1990)\citenamefont
  {Goderis}, \citenamefont {Verbeure},\ and\ \citenamefont
  {Vets}}]{goderis1990}%
  \BibitemOpen
  \bibfield  {author} {\bibinfo {author} {\bibfnamefont {D.}~\bibnamefont
  {Goderis}}, \bibinfo {author} {\bibfnamefont {A.}~\bibnamefont {Verbeure}},\
  and\ \bibinfo {author} {\bibfnamefont {P.}~\bibnamefont {Vets}},\ }\bibfield
  {title} {\bibinfo {title} {Dynamics of fluctuations for quantum lattice
  systems},\ }\href {https://doi.org/10.1007/BF02096872} {\bibfield  {journal}
  {\bibinfo  {journal} {Commun. Math. Phys.}\ }\textbf {\bibinfo {volume}
  {128}},\ \bibinfo {pages} {533} (\bibinfo {year} {1990})}\BibitemShut
  {NoStop}%
\bibitem [{\citenamefont {Benatti}\ \emph {et~al.}(2017)\citenamefont
  {Benatti}, \citenamefont {Carollo}, \citenamefont {Floreanini},\ and\
  \citenamefont {Narnhofer}}]{benatti2017}%
  \BibitemOpen
  \bibfield  {author} {\bibinfo {author} {\bibfnamefont {F.}~\bibnamefont
  {Benatti}}, \bibinfo {author} {\bibfnamefont {F.}~\bibnamefont {Carollo}},
  \bibinfo {author} {\bibfnamefont {R.}~\bibnamefont {Floreanini}},\ and\
  \bibinfo {author} {\bibfnamefont {H.}~\bibnamefont {Narnhofer}},\ }\bibfield
  {title} {\bibinfo {title} {Quantum fluctuations in mesoscopic systems},\
  }\href {https://doi.org/10.1088/1751-8121/aa84d2} {\bibfield  {journal}
  {\bibinfo  {journal} {J. Phys. A: Math. Theor.}\ }\textbf {\bibinfo {volume}
  {50}},\ \bibinfo {pages} {423001} (\bibinfo {year} {2017})}\BibitemShut
  {NoStop}%
\bibitem [{\citenamefont {Stitely}\ \emph {et~al.}(2023)\citenamefont
  {Stitely}, \citenamefont {Finger}, \citenamefont {Rosa-Medina}, \citenamefont
  {Ferri}, \citenamefont {Donner}, \citenamefont {Esslinger}, \citenamefont
  {Parkins},\ and\ \citenamefont {Krauskopf}}]{stitely2023}%
  \BibitemOpen
  \bibfield  {author} {\bibinfo {author} {\bibfnamefont {K.}~\bibnamefont
  {Stitely}}, \bibinfo {author} {\bibfnamefont {F.}~\bibnamefont {Finger}},
  \bibinfo {author} {\bibfnamefont {R.}~\bibnamefont {Rosa-Medina}}, \bibinfo
  {author} {\bibfnamefont {F.}~\bibnamefont {Ferri}}, \bibinfo {author}
  {\bibfnamefont {T.}~\bibnamefont {Donner}}, \bibinfo {author} {\bibfnamefont
  {T.}~\bibnamefont {Esslinger}}, \bibinfo {author} {\bibfnamefont
  {S.}~\bibnamefont {Parkins}},\ and\ \bibinfo {author} {\bibfnamefont
  {B.}~\bibnamefont {Krauskopf}},\ }\bibfield  {title} {\bibinfo {title}
  {{Quantum Fluctuation Dynamics of Dispersive Superradiant Pulses in a Hybrid
  Light-Matter System}},\ }\href {https://doi.org/10.48550/arXiv.2302.08078}
  {\bibfield  {journal} {\bibinfo  {journal} {arXiv:2302.08078}\ } (\bibinfo
  {year} {2023})}\BibitemShut {NoStop}%
\bibitem [{\citenamefont {Lindblad}(1976)}]{lindblad1976}%
  \BibitemOpen
  \bibfield  {author} {\bibinfo {author} {\bibfnamefont {G.}~\bibnamefont
  {Lindblad}},\ }\bibfield  {title} {\bibinfo {title} {{On the generators of
  quantum dynamical semigroups}},\ }\href {https://doi.org/10.1007/BF01608499}
  {\bibfield  {journal} {\bibinfo  {journal} {Commun. Math. Phys.}\ }\textbf
  {\bibinfo {volume} {48}},\ \bibinfo {pages} {119} (\bibinfo {year}
  {1976})}\BibitemShut {NoStop}%
\bibitem [{\citenamefont {Gorini}\ \emph {et~al.}(1976)\citenamefont {Gorini},
  \citenamefont {Kossakowski},\ and\ \citenamefont {Sudarshan}}]{gorini1976}%
  \BibitemOpen
  \bibfield  {author} {\bibinfo {author} {\bibfnamefont {V.}~\bibnamefont
  {Gorini}}, \bibinfo {author} {\bibfnamefont {A.}~\bibnamefont
  {Kossakowski}},\ and\ \bibinfo {author} {\bibfnamefont {E.~C.~G.}\
  \bibnamefont {Sudarshan}},\ }\bibfield  {title} {\bibinfo {title}
  {{Completely positive dynamical semigroups of N‐level systems}},\ }\href
  {https://doi.org/10.1063/1.522979} {\bibfield  {journal} {\bibinfo  {journal}
  {J. Math. Phys.}\ }\textbf {\bibinfo {volume} {17}},\ \bibinfo {pages} {821}
  (\bibinfo {year} {1976})}\BibitemShut {NoStop}%
\bibitem [{\citenamefont {Breuer}\ and\ \citenamefont
  {Petruccione}(2002)}]{breuer2002theory}%
  \BibitemOpen
  \bibfield  {author} {\bibinfo {author} {\bibfnamefont {H.-P.}\ \bibnamefont
  {Breuer}}\ and\ \bibinfo {author} {\bibfnamefont {F.}~\bibnamefont
  {Petruccione}},\ }\href@noop {} {\emph {\bibinfo {title} {{The theory of open
  quantum systems}}}}\ (\bibinfo  {publisher} {Oxford University Press on
  Demand},\ \bibinfo {year} {2002})\BibitemShut {NoStop}%
\bibitem [{\citenamefont {Tavis}\ and\ \citenamefont
  {Cummings}(1967)}]{tavis1967}%
  \BibitemOpen
  \bibfield  {author} {\bibinfo {author} {\bibfnamefont {M.}~\bibnamefont
  {Tavis}}\ and\ \bibinfo {author} {\bibfnamefont {F.}~\bibnamefont
  {Cummings}},\ }\bibfield  {title} {\bibinfo {title} {{The exact solution of N
  two level systems interacting with a single mode, quantized radiation
  field}},\ }\href
  {https://doi.org/https://doi.org/10.1016/0375-9601(67)90957-7} {\bibfield
  {journal} {\bibinfo  {journal} {Phys. Lett. A}\ }\textbf {\bibinfo {volume}
  {25}},\ \bibinfo {pages} {714} (\bibinfo {year} {1967})}\BibitemShut
  {NoStop}%
\bibitem [{\citenamefont {Tavis}\ and\ \citenamefont
  {Cummings}(1969)}]{tavis1969}%
  \BibitemOpen
  \bibfield  {author} {\bibinfo {author} {\bibfnamefont {M.}~\bibnamefont
  {Tavis}}\ and\ \bibinfo {author} {\bibfnamefont {F.~W.}\ \bibnamefont
  {Cummings}},\ }\bibfield  {title} {\bibinfo {title} {{Approximate Solutions
  for an $N$-Molecule-Radiation-Field Hamiltonian}},\ }\href
  {https://doi.org/10.1103/PhysRev.188.692} {\bibfield  {journal} {\bibinfo
  {journal} {Phys. Rev.}\ }\textbf {\bibinfo {volume} {188}},\ \bibinfo {pages}
  {692} (\bibinfo {year} {1969})}\BibitemShut {NoStop}%
\bibitem [{SM()}]{SM}%
  \BibitemOpen
  \href@noop {} {}\bibinfo {note} {See Supplemental Material, which further
  contains
  Refs.~\cite{benatti2015,verbeure2010,bratteli1982,thirring2013,strocchi2021},
  for details on the proofs of the main results.}\BibitemShut {Stop}%
\bibitem [{\citenamefont {Bratteli}\ and\ \citenamefont
  {Robinson}(1981)}]{bratteli1982}%
  \BibitemOpen
  \bibfield  {author} {\bibinfo {author} {\bibfnamefont {O.}~\bibnamefont
  {Bratteli}}\ and\ \bibinfo {author} {\bibfnamefont {D.~W.}\ \bibnamefont
  {Robinson}},\ }\href@noop {} {\emph {\bibinfo {title} {{Operator Algebras and
  Quantum Statistical Mechanics II. Equilibrium States Models in Quantum
  Statistical Mechanics}}}}\ (\bibinfo  {publisher} {Springer Berlin,
  Heidelberg},\ \bibinfo {year} {1981})\BibitemShut {NoStop}%
\bibitem [{\citenamefont {Thirring}(2013)}]{thirring2013}%
  \BibitemOpen
  \bibfield  {author} {\bibinfo {author} {\bibfnamefont {W.}~\bibnamefont
  {Thirring}},\ }\href@noop {} {\emph {\bibinfo {title} {Quantum mathematical
  physics: atoms, molecules and large systems}}}\ (\bibinfo  {publisher}
  {Springer Science \& Business Media},\ \bibinfo {year} {2013})\BibitemShut
  {NoStop}%
\bibitem [{\citenamefont {Strocchi}(2021)}]{strocchi2021}%
  \BibitemOpen
  \bibfield  {author} {\bibinfo {author} {\bibfnamefont {F.}~\bibnamefont
  {Strocchi}},\ }\href@noop {} {\emph {\bibinfo {title} {Symmetry breaking}}}\
  (\bibinfo  {publisher} {Springer Berlin, Heidelberg},\ \bibinfo {year}
  {2021})\BibitemShut {NoStop}%
\bibitem [{\citenamefont {Verbeure}(2010)}]{verbeure2010}%
  \BibitemOpen
  \bibfield  {author} {\bibinfo {author} {\bibfnamefont {A.~F.}\ \bibnamefont
  {Verbeure}},\ }\href@noop {} {\emph {\bibinfo {title} {Many-body boson
  systems: half a century later}}}\ (\bibinfo  {publisher} {Springer},\
  \bibinfo {year} {2010})\BibitemShut {NoStop}%
\bibitem [{\citenamefont {Benatti}\ \emph {et~al.}(2015)\citenamefont
  {Benatti}, \citenamefont {Carollo},\ and\ \citenamefont
  {Floreanini}}]{benatti2015}%
  \BibitemOpen
  \bibfield  {author} {\bibinfo {author} {\bibfnamefont {F.}~\bibnamefont
  {Benatti}}, \bibinfo {author} {\bibfnamefont {F.}~\bibnamefont {Carollo}},\
  and\ \bibinfo {author} {\bibfnamefont {R.}~\bibnamefont {Floreanini}},\
  }\bibfield  {title} {\bibinfo {title} {{Dissipative dynamics of quantum
  fluctuations}},\ }\href
  {https://doi.org/https://doi.org/10.1002/andp.201500165} {\bibfield
  {journal} {\bibinfo  {journal} {Ann. Phys. (Berl.)}\ }\textbf {\bibinfo
  {volume} {527}},\ \bibinfo {pages} {639} (\bibinfo {year}
  {2015})}\BibitemShut {NoStop}%
\bibitem [{\citenamefont {Narnhofer}\ and\ \citenamefont
  {Thirring}(2002)}]{narnhofer2002}%
  \BibitemOpen
  \bibfield  {author} {\bibinfo {author} {\bibfnamefont {H.}~\bibnamefont
  {Narnhofer}}\ and\ \bibinfo {author} {\bibfnamefont {W.}~\bibnamefont
  {Thirring}},\ }\bibfield  {title} {\bibinfo {title} {{Entanglement of
  mesoscopic systems}},\ }\href {https://doi.org/10.1103/PhysRevA.66.052304}
  {\bibfield  {journal} {\bibinfo  {journal} {Phys. Rev. A}\ }\textbf {\bibinfo
  {volume} {66}},\ \bibinfo {pages} {052304} (\bibinfo {year}
  {2002})}\BibitemShut {NoStop}%
\bibitem [{\citenamefont {Carollo}\ and\ \citenamefont
  {Lesanovsky}(2022)}]{carollo2022}%
  \BibitemOpen
  \bibfield  {author} {\bibinfo {author} {\bibfnamefont {F.}~\bibnamefont
  {Carollo}}\ and\ \bibinfo {author} {\bibfnamefont {I.}~\bibnamefont
  {Lesanovsky}},\ }\bibfield  {title} {\bibinfo {title} {{Exact solution of a
  boundary time-crystal phase transition: Time-translation symmetry breaking
  and non-Markovian dynamics of correlations}},\ }\href
  {https://doi.org/10.1103/PhysRevA.105.L040202} {\bibfield  {journal}
  {\bibinfo  {journal} {Phys. Rev. A}\ }\textbf {\bibinfo {volume} {105}},\
  \bibinfo {pages} {L040202} (\bibinfo {year} {2022})}\BibitemShut {NoStop}%
\bibitem [{\citenamefont {Matsui}(2003)}]{matsui2003}%
  \BibitemOpen
  \bibfield  {author} {\bibinfo {author} {\bibfnamefont {T.}~\bibnamefont
  {Matsui}},\ }\bibfield  {title} {\bibinfo {title} {{On the Algebra of
  Fluctuation in Quantum Spin Chains}},\ }\href
  {https://doi.org/10.1007/s00023-003-0122-z} {\bibfield  {journal} {\bibinfo
  {journal} {Ann. Henri Poincar{\'e}}\ }\textbf {\bibinfo {volume} {4}},\
  \bibinfo {pages} {63} (\bibinfo {year} {2003})}\BibitemShut {NoStop}%
\bibitem [{\citenamefont {Lanford}\ and\ \citenamefont
  {Ruelle}(1969)}]{landford1969}%
  \BibitemOpen
  \bibfield  {author} {\bibinfo {author} {\bibfnamefont {O.~E.}\ \bibnamefont
  {Lanford}}\ and\ \bibinfo {author} {\bibfnamefont {D.}~\bibnamefont
  {Ruelle}},\ }\bibfield  {title} {\bibinfo {title} {{Observables at infinity
  and states with short range correlations in statistical mechanics}},\ }\href
  {https://doi.org/10.1007/BF01645487} {\bibfield  {journal} {\bibinfo
  {journal} {Commun. Math. Phys.}\ }\textbf {\bibinfo {volume} {13}},\ \bibinfo
  {pages} {194} (\bibinfo {year} {1969})}\BibitemShut {NoStop}%
\bibitem [{\citenamefont {Davies}(1973)}]{davies1973}%
  \BibitemOpen
  \bibfield  {author} {\bibinfo {author} {\bibfnamefont {E.~B.}\ \bibnamefont
  {Davies}},\ }\bibfield  {title} {\bibinfo {title} {{Exact dynamics of an
  infinite-atom Dicke maser model}},\ }\href
  {https://doi.org/10.1007/BF01667916} {\bibfield  {journal} {\bibinfo
  {journal} {Commun. Math. Phys.}\ }\textbf {\bibinfo {volume} {33}},\ \bibinfo
  {pages} {187} (\bibinfo {year} {1973})}\BibitemShut {NoStop}%
\bibitem [{\citenamefont {Rovnyak}(2008)}]{rovnyak2008}%
  \BibitemOpen
  \bibfield  {author} {\bibinfo {author} {\bibfnamefont {D.}~\bibnamefont
  {Rovnyak}},\ }\bibfield  {title} {\bibinfo {title} {{Tutorial on analytic
  theory for cross-polarization in solid state NMR}},\ }\href
  {https://doi.org/https://doi.org/10.1002/cmr.a.20115} {\bibfield  {journal}
  {\bibinfo  {journal} {Conc. Magnet. Reson. A}\ }\textbf {\bibinfo {volume}
  {32A}},\ \bibinfo {pages} {254} (\bibinfo {year} {2008})}\BibitemShut
  {NoStop}%
\bibitem [{\citenamefont {Holstein}\ and\ \citenamefont
  {Primakoff}(1940)}]{holstein1940}%
  \BibitemOpen
  \bibfield  {author} {\bibinfo {author} {\bibfnamefont {T.}~\bibnamefont
  {Holstein}}\ and\ \bibinfo {author} {\bibfnamefont {H.}~\bibnamefont
  {Primakoff}},\ }\bibfield  {title} {\bibinfo {title} {{Field Dependence of
  the Intrinsic Domain Magnetization of a Ferromagnet}},\ }\href
  {https://doi.org/10.1103/PhysRev.58.1098} {\bibfield  {journal} {\bibinfo
  {journal} {Phys. Rev.}\ }\textbf {\bibinfo {volume} {58}},\ \bibinfo {pages}
  {1098} (\bibinfo {year} {1940})}\BibitemShut {NoStop}%
\bibitem [{\citenamefont {Yuan}\ \emph {et~al.}(2007)\citenamefont {Yuan},
  \citenamefont {Goan},\ and\ \citenamefont {Zhu}}]{yuan2007}%
  \BibitemOpen
  \bibfield  {author} {\bibinfo {author} {\bibfnamefont {X.-Z.}\ \bibnamefont
  {Yuan}}, \bibinfo {author} {\bibfnamefont {H.-S.}\ \bibnamefont {Goan}},\
  and\ \bibinfo {author} {\bibfnamefont {K.-D.}\ \bibnamefont {Zhu}},\
  }\bibfield  {title} {\bibinfo {title} {Non-markovian reduced dynamics and
  entanglement evolution of two coupled spins in a quantum spin environment},\
  }\href {https://doi.org/10.1103/PhysRevB.75.045331} {\bibfield  {journal}
  {\bibinfo  {journal} {Phys. Rev. B}\ }\textbf {\bibinfo {volume} {75}},\
  \bibinfo {pages} {045331} (\bibinfo {year} {2007})}\BibitemShut {NoStop}%
\bibitem [{\citenamefont {Dehghani}\ \emph {et~al.}(2020)\citenamefont
  {Dehghani}, \citenamefont {Mojaveri},\ and\ \citenamefont
  {Vaez}}]{dehghani2020}%
  \BibitemOpen
  \bibfield  {author} {\bibinfo {author} {\bibfnamefont {A.}~\bibnamefont
  {Dehghani}}, \bibinfo {author} {\bibfnamefont {B.}~\bibnamefont {Mojaveri}},\
  and\ \bibinfo {author} {\bibfnamefont {M.}~\bibnamefont {Vaez}},\ }\bibfield
  {title} {\bibinfo {title} {Entanglement dynamics of two coupled spins
  interacting with an adjustable spin bath: effect of an exponential variable
  magnetic field},\ }\href {https://doi.org/10.1007/s11128-020-02803-5}
  {\bibfield  {journal} {\bibinfo  {journal} {Quantum Inf. Process.}\ }\textbf
  {\bibinfo {volume} {19}},\ \bibinfo {pages} {306} (\bibinfo {year}
  {2020})}\BibitemShut {NoStop}%
\bibitem [{\citenamefont {Ashida}\ \emph {et~al.}(2019)\citenamefont {Ashida},
  \citenamefont {Shi}, \citenamefont {Schmidt}, \citenamefont {Sadeghpour},
  \citenamefont {Cirac},\ and\ \citenamefont {Demler}}]{ashida2019}%
  \BibitemOpen
  \bibfield  {author} {\bibinfo {author} {\bibfnamefont {Y.}~\bibnamefont
  {Ashida}}, \bibinfo {author} {\bibfnamefont {T.}~\bibnamefont {Shi}},
  \bibinfo {author} {\bibfnamefont {R.}~\bibnamefont {Schmidt}}, \bibinfo
  {author} {\bibfnamefont {H.~R.}\ \bibnamefont {Sadeghpour}}, \bibinfo
  {author} {\bibfnamefont {J.~I.}\ \bibnamefont {Cirac}},\ and\ \bibinfo
  {author} {\bibfnamefont {E.}~\bibnamefont {Demler}},\ }\bibfield  {title}
  {\bibinfo {title} {{Quantum Rydberg Central Spin Model}},\ }\href
  {https://doi.org/10.1103/PhysRevLett.123.183001} {\bibfield  {journal}
  {\bibinfo  {journal} {Phys. Rev. Lett.}\ }\textbf {\bibinfo {volume} {123}},\
  \bibinfo {pages} {183001} (\bibinfo {year} {2019})}\BibitemShut {NoStop}%
\bibitem [{\citenamefont {Anikeeva}\ \emph {et~al.}(2021)\citenamefont
  {Anikeeva}, \citenamefont {Markovi\ifmmode~\acute{c}\else \'{c}\fi{}},
  \citenamefont {Borish}, \citenamefont {Hines}, \citenamefont {Rajagopal},
  \citenamefont {Cooper}, \citenamefont {Periwal}, \citenamefont
  {Safavi-Naeini}, \citenamefont {Davis},\ and\ \citenamefont
  {Schleier-Smith}}]{anikeeva2021}%
  \BibitemOpen
  \bibfield  {author} {\bibinfo {author} {\bibfnamefont {G.}~\bibnamefont
  {Anikeeva}}, \bibinfo {author} {\bibfnamefont {O.}~\bibnamefont
  {Markovi\ifmmode~\acute{c}\else \'{c}\fi{}}}, \bibinfo {author}
  {\bibfnamefont {V.}~\bibnamefont {Borish}}, \bibinfo {author} {\bibfnamefont
  {J.~A.}\ \bibnamefont {Hines}}, \bibinfo {author} {\bibfnamefont {S.~V.}\
  \bibnamefont {Rajagopal}}, \bibinfo {author} {\bibfnamefont {E.~S.}\
  \bibnamefont {Cooper}}, \bibinfo {author} {\bibfnamefont {A.}~\bibnamefont
  {Periwal}}, \bibinfo {author} {\bibfnamefont {A.}~\bibnamefont
  {Safavi-Naeini}}, \bibinfo {author} {\bibfnamefont {E.~J.}\ \bibnamefont
  {Davis}},\ and\ \bibinfo {author} {\bibfnamefont {M.}~\bibnamefont
  {Schleier-Smith}},\ }\bibfield  {title} {\bibinfo {title} {{Number
  Partitioning With Grover's Algorithm in Central Spin Systems}},\ }\href
  {https://doi.org/10.1103/PRXQuantum.2.020319} {\bibfield  {journal} {\bibinfo
   {journal} {PRX Quantum}\ }\textbf {\bibinfo {volume} {2}},\ \bibinfo {pages}
  {020319} (\bibinfo {year} {2021})}\BibitemShut {NoStop}%
\bibitem [{\citenamefont {Dobrzyniecki}\ and\ \citenamefont
  {Tomza}(2023)}]{dobrzyniecki2023}%
  \BibitemOpen
  \bibfield  {author} {\bibinfo {author} {\bibfnamefont {J.}~\bibnamefont
  {Dobrzyniecki}}\ and\ \bibinfo {author} {\bibfnamefont {M.}~\bibnamefont
  {Tomza}},\ }\bibfield  {title} {\bibinfo {title} {{Quantum simulation of the
  central spin model with a Rydberg atom and polar molecules in optical
  tweezers}},\ }\href {https://doi.org/10.48550/arXiv.2302.14774} {\bibfield
  {journal} {\bibinfo  {journal} {arXiv:2302.14774}\ } (\bibinfo {year}
  {2023})}\BibitemShut {NoStop}%
\end{thebibliography}%
\newpage
\setcounter{equation}{0}
\setcounter{figure}{0}
\setcounter{table}{0}
\makeatletter
\renewcommand{\theequation}{S\arabic{equation}}
\renewcommand{\thefigure}{S\arabic{figure}}
\makeatletter

\onecolumngrid
\newpage

\setcounter{page}{1}
\begin{center}
{\Large SUPPLEMENTAL MATERIAL}
\end{center}
\begin{center}
\vspace{0.8cm}
{\Large Non-Gaussian dynamics of quantum fluctuations and mean-field limit in open quantum central spin systems}
\end{center}
\begin{center}
Federico Carollo \\
{\em Institut f\"ur Theoretische Physik, Universit\"at T\"ubingen,}\\
{\em Auf der Morgenstelle 14, 72076 T\"ubingen, Germany}\\

\end{center}

\section*{I. Behavior of local bath-spin operators}
In this Section, we first give a definition of  local bath-spin operators for the central spin system and then prove Lemma \ref{local-bathspin}. 
For mathematical convenience, we consider the bath-spin system to be infinite and define, as usually done, the dynamical generators as involving only $N$ bath spins (see main text). We then analyze their behavior in the limit $N\to\infty$. For the (infinite) bath-spin system, the reference algebra is the so-called quasi-local algebra $\mathcal{A}$ \cite{,bratteli1982,thirring2013,strocchi2021}, which contains all operator sequences converging in norm. A bath-spin operator $A\in \mathcal{A}$ is said to be (strictly) local if it has finite support $\Lambda_A$ or, in other words, if it acts in a nontrivial way only on a finite number of bath spins. The support of $A$ can be defined as the set
$$
\Lambda_A=\left\{k\in \mathbb{N}^+: \sum_{\alpha=x,y,z}\left\|[\sigma_\alpha^{(k)},A]\right\|>0 \right\}\, .
$$

{\it Proof of Lemma \ref{local-bathspin}:} Let us consider the difference between the evolution of a local bath-spin operator $A$ as implemented by the generator $\mathcal{L}$ and as implemented by the generator $\mathcal{D}_{\rm bath}$. We can write such a difference as 
$$
e^{t\mathcal{L}}[A]-e^{t\mathcal{D}_{\rm bath}}[A]=\int_0^t {\rm d}s \frac{\rm d }{{\rm d}s} \left(e^{s\mathcal{L}}\circ e^{(t-s)\mathcal{D}_{\rm bath}}[A]\right)=\int_0^t {\rm d}s \, e^{s\mathcal{L}}\left[i\left[H_{\rm int},e^{(t-s)\mathcal{D}_{\rm bath}}[A]\right]\right]\, ,
$$ 
where we used that $e^{(t-s)\mathcal{D}_{\rm bath}}[A]$ is a bath-spin operator to get rid of the maps acting on the central spin only. Moreover, since $\mathcal{D}_{\rm bath}$ acts independently on bath spins, $e^{(t-s)\mathcal{D}_{\rm bath}}[A]$ has the same support as $A$. Exploiting this observation and considering, since we are interested in the limit $N\to\infty$, $N\ge\max(\Lambda_A)$, we find that 
$$
i\left[H_{\rm int}, e^{(t-s)\mathcal{D}_{\rm bath}}[A]\right]=ig\tau_+\sum_{k\in\Lambda_A} \left[\sigma_-^{(k)},e^{(t-s)\mathcal{D}_{\rm bath}}[A]\right]+ig \tau_-\sum_{k\in\Lambda_A} \left[\sigma_+^{(k)}, e^{(t-s)\mathcal{D}_{\rm bath}}[A]\right]\, .
$$
We further note that $\left\|e^{(t-s)\mathcal{D_{\rm bath}}}[O]\right\|\le \|O\|$ for any operator $O$ since $ \mathcal{D}_{\rm bath}$ implements a contractive map. The same is  true for $\mathcal{L}$. We can thus write that 
$$
\left\|e^{s\mathcal{L}}\left[i\left[H_{\rm int},e^{(t-s)\mathcal{D}_{\rm bath}}[A]\right]\right]\right\|\le \left\|\left[H_{\rm int}, e^{(t-s)\mathcal{D}_{\rm bath}}[A]\right]\right\|\le 4|g|\left|\Lambda_A\right|\|A\|\, ,
$$
where $\left|\Lambda_A\right|$ denotes the cardinality of the set $\Lambda_A$. Substituting the assumed form of $g$, i.e., $g=g_0/N^{z}$, we find 
$$
\left\|e^{t\mathcal{L}}[A]-e^{t\mathcal{D}_{\rm bath}}[A]\right\|\le \frac{4|g_0|t}{N^z}\left|\Lambda_A\right| \|A\|\, ,
$$
which vanishes, in the thermodynamic limit, whenever $z>0$. \qed
\\

The above result extends to the average operators $m_\alpha^N=\sum_{k=1}^N\sigma_\alpha^{(k)}/N$ considered in the main text. This is achieved by exploiting the linearity of the maps $\mathcal{D}_{\rm bath}$ and $\mathcal{L}$. 
We indeed have 
$$
\left\|e^{t\mathcal{L}}[m_\alpha^N]-e^{t\mathcal{D}_{\rm bath}}[m_\alpha^N]\right\| \le \frac{1}{N}\sum_{k=1}^N \left\|e^{t\mathcal{L}}[\sigma_\alpha^{(k)}]-e^{t\mathcal{D}_{\rm bath}}[\sigma_\alpha^{(k)}]\right\|\le \frac{4|g_0|t}{N^{z}}\, , 
$$
where we have used Lemma \ref{local-bathspin} and  that $\left|\Lambda_{\sigma_\alpha^{(k)}}\right|=1$, $\forall k$. 

\section*{II. Emergent bosonic subsystem}
In the following, we provide a proof of Proposition \ref{QCLT}, which establishes the convergence, in a quantum central limit sense, of the finite-$N$ operators $a_N,a_N^\dagger$ to bosonic operators. This convergence is rather a mapping of the bath-spin system onto a bosonic one \cite{verbeure2010}. For the purpose of the proof, we introduce the exponential (displacement-like) operators 
$D_N(s)=e^{sa_N^\dagger -s^* a_N}$. The aim is to interpret any possible expectation value of these operators (and their products) on the state $\Omega_{\rm SS}$, as an expectation value of proper displacement operators $D(s)=e^{s a^\dagger -s^*a}$ on a Gaussian state. This interpretation necessarily also defines the state $\tilde{\Omega}_\beta$ on the bosonic operators $D(s)$, as it emerges from $\Omega_{\rm SS}$. \\

{\it Proof of Proposition \ref{QCLT}:} We start by calculating $\Omega_{\rm SS }\left(D_N(s)\right)$. Exploiting the definition of $a_N,a_N^\dagger$ and using the uncorrelated structure of the state, we have
$$
\Omega_{\rm SS }\left(D_N(s)\right)=\prod_{k=1}^N \Omega_{\rm SS}\left(\exp\left(\frac{s \sigma_+^{(k)}}{\sqrt{\varepsilon N}}-\frac{s^* \sigma_-^{(k)}}{\sqrt{\varepsilon N}}\right)\right) =\left[\Omega_{\rm SS}\left(\exp\left(\frac{s \sigma_+^{(k)}}{\sqrt{\varepsilon N}}-\frac{s^* \sigma_-^{(k)}}{\sqrt{\varepsilon N}}\right)\right)\right]^N\, ,
$$
where we have also exploited the permutation invariance of the state. Expanding the exponential, we find (we remove the label $k$ since the quantity does not depend on $k$) 
$$
\Omega_{\rm SS }\left(D_N(s)\right)\approx\left[1+\Omega_{\rm SS}\left(\frac{\left(s\sigma_+-s^*\sigma_-\right)^2}{2 \varepsilon N}\right)\right]^N=\left(1-\frac{|s|^2}{2\varepsilon N}\right)^N\, .
$$
Taking the large-$N$ limit, we find $
\lim_{N\to\infty}\Omega_{\rm SS}\left(D_N(s)\right)=e^{-\frac{|s^2|}{2\varepsilon }}
$.\\

We now perform the analogous calculation for the bosonic displacement operator. Considering $\tilde{\Omega}_\beta$, we find
$$
\tilde{\Omega}_\beta\left(D(s)\right)=\sum_{k=0}^\infty\frac{1}{k!}\tilde{\Omega}_\beta \left((sa^\dagger -s^* a)^k\right)=\sum_{k=0}^\infty \frac{1}{(2k)!}\tilde{\Omega}_\beta \left((sa^\dagger -s^* a)^{2k}\right)\, ,
$$
where we used that the thermal state $\tilde{\Omega}_\beta$ is diagonal in the number-operator basis and thus only terms with even power  are non-vanishing. The thermal state is Gaussian and,  exploiting Isserlis theorem, we can write 
$$
\tilde{\Omega}_\beta\left(D(s)\right)=\sum_{k=0}^\infty \frac{1}{2^k k!}C^k =e^{\frac{C}{2}}\, ,
$$
where $C=\tilde{\Omega}_\beta \left((sa^\dagger -s^* a)^{2}\right)$ is given by 
$
C=-|s|^2 (1+2n_\beta)=-\frac{|s|^2}{\varepsilon}
$, due to our choice of $\beta\omega$. This demonstrates that 
$
\lim_{N\to\infty}\Omega_{\rm SS}\left(D_N(s)\right)=\tilde{\Omega}_\beta\left(D(s)\right)
$.

Similar results are valid for arbitrary products of displacement-like operators $D_N(s_i)$. We start showing this for a product of two displacement-like operators. 
We thus consider the expectation $\Omega_{\rm SS}\left(D_N(s_1) D_N(s_2)\right)$ and the first task is to combine the two exponentials. Using Baker-Campbell-Hausdorff, together with, e.g., Remark 3 in Ref.~\cite{benatti2015} or Eq.~(6.18) in Ref.~\cite{verbeure2010}, we can write
\begin{equation}
\label{composition}
D_N(s_1)D_N(s_2)= D_N(s_1+s_2)e^{\frac{-m_z^N}{2\varepsilon}\left(s_1s_2^*-s_1^*s_2\right)}+O\left(\frac{1}{N}\right)\, .
\end{equation}
Using that $m_z^N$ is an average operator which converges to a multiple of the identity on clustering states, such as $\Omega_{\rm SS}$, and noticing that $\Omega_{\rm SS}(m_z^N)=-\varepsilon$, we have that 
\begin{equation*}
\lim_{N\to\infty}\Omega_{\rm SS}\left(D_N(s_1)D_N(s_2)\right)=e^{\frac{s_1s_2^*-s_1^*s_2}{2}}\lim_{N\to\infty}\Omega_{\rm SS}\left(D_N(s_1+s_2)\right)=e^{\frac{s_1s_2^*-s_1^*s_2}{2}}e^{-\frac{|s_1+s_2|^2}{2\varepsilon}}\, .
\end{equation*}
Exploiting the Baker-Campbell-Hausdorff formula for bosonic displacement operators and the result for the expectation of a single displacement operator obtained before, we have
$$
\lim_{N\to\infty }\Omega_{\rm SS}\left(D_N(s_1)D_N(s_2)\right)=\tilde{\Omega}_\beta\left(D(s_1)D(s_2)\right)\, .
$$

To extend this result to products of three displacement-like operators $D_N(s_1)D_N(s_2)D_N(s_3)$ one can use the composition rule in Eq.~\eqref{composition} for, e.g., $D_N(s_2)D_N(s_3)$ and account for the correction. This then reduces to the case of a product of two displacement-like operators $D_N(s_1)D_N(s_2+s_3)$, already discussed. The argument can be extended to the product of an  arbitrary number of displacement-like operators by induction. \qed

\section*{III. Dynamical generator for the one-spin One-boson system}
In this Section, we prove Theorem \ref{Theo1} establishing the form of the  generator for the emergent one-spin one-boson system. We later show that [see Corollary \ref{Cor1}], assuming reasonable conditions on the dynamics (expected to hold from a physical perspective in all practical cases), Theorem \ref{Theo1} implies that the dynamics of the central spin system is captured, in the thermodynamic limit, by the one-spin one-boson model evolving through the generator in Eq.~\eqref{L_tilde}. \\

In preparation to the proof of Theorem \ref{Theo1}, we introduce a class of monomials involving products of the relevant operators, i.e., operators of the central spin, powers of fluctuation operators and of the operator $m_z^N$. We  define
\begin{equation}
P_N^{\vec{\mu}}=\tau_{\mu_1} a_N^{\dagger\, \mu_2}a_N^{\mu_3}m_z^{N\, \mu_4}\, ,
\label{poly}
\end{equation}
where $\tau_{\mu_1}$ is a Pauli matrix of the central spin, while $\mu_2,\mu_3,\mu_4\in \mathbb{N}$ specify the powers for the remaining operators. We recall already here that, in the state $\Omega_{\rm SS}$, $m_z^N$ tends to $-\varepsilon$. The monomials $P_N^{\vec{\mu}}$ also contain operators acting on the central spin. In order to calculate the full expectation value of these monomials, we thus have to introduce a state $\varphi(\cdot)$ for the central spin. We thus write the (initial) state of the central spin system as $\varphi\otimes \Omega_{\rm SS}$. 

Due to Proposition \ref{QCLT}, the expectation value of monomials of fluctuation operators converges to the analogous expectation constructed in terms of the bosonic operators $a,a^\dagger$ over the state $\tilde{\Omega}_\beta$. This means that 
$$
\lim_{N\to\infty} \varphi\otimes \Omega_{\rm SS}\left(P_N^{\vec{\nu}}P_N^{\vec{\mu}}P_N^{\vec{\eta}}\right)=\varphi\otimes\tilde{\Omega}_\beta\left(P^{\vec{\nu}}P^{\vec{\mu}}P^{\vec{\eta}}\right)=\varphi\left(\tau_{\nu_1}\tau_{\mu_1}\tau_{\eta_1}\right) \tilde{\Omega}_{\beta}\left(a^{\dagger\, \nu_2}a^{\nu_3}a^{\dagger\, \mu_2}a^{\mu_3}a^{\dagger\, \eta_2}a^{\eta_3}\right)(-\varepsilon)^{\nu_4+\mu_4+\eta_4}\, ,
$$
where we have also defined the monomial $
P^{\vec{\mu}}=\tau_{\mu_1} a^{\dagger\, \mu_2}a^{\mu_3}\left(-\varepsilon\right)^{\mu_4}
$.
The above limit specifies in which sense $P_N^{\vec{\mu}}$ tends to the monomial $P^{\vec{\mu}}$. We further consider the product of three monomials because this allows us to control all the possible ``matrix elements" of the monomial in the middle by varying the monomials on both sides, just as one would do in a standard weak-operator topology \cite{bratteli1982,thirring2013,strocchi2021}. We also note  that the monomials $P_N^{\vec{\mu}}$ define equivalence classes since, for instance, the monomial  $P_N^{\vec{\mu}}+O(1/N)$, where $O(1/N)$ indicates a quantity that vanishes as $1/N$ under the expectation written above, still converges to the monomial $P^{\vec{\mu}}$.

The structure of the above limit is thus important to understand the action of the generator on generic monomials, ${\mathcal{L}}[P_N^{\vec{\mu}}]$, by controlling all of its possible matrix elements, or physically speaking all correlation functions. In fact, to prove the Theorem, we want to demonstrate that the generator $\tilde{\mathcal{L}}$, as defined in the main text, is such that 
\begin{equation}
\label{limit-generator}
\lim_{N\to\infty}\varphi\otimes \Omega_{\rm SS}\left(P_N^{\vec{\nu}}{\mathcal{L}}\left[P_N^{\vec{\mu}}\right]P_N^{\vec{\eta}}\right)=\varphi\otimes \tilde{\Omega}_\beta\left(P^{\vec{\nu}}{\tilde{\mathcal{L}}}\left[P^{\vec{\mu}}\right]P^{\vec{\eta}}\right)\, ,\qquad  \forall \vec{\nu},\vec{\mu},\vec{\eta}\, .
\end{equation}
This means that the generator $\tilde{\mathcal{L}}$ acting on the emergent algebra is able to reproduce the action of $\mathcal{L}$ in the thermodynamic limit. \\

{\it Proof of Theorem \ref{Theo1}:} We start by analyzing the action of the generator $\mathcal{L}$ on a generic monomial. We have that 
$$
\mathcal{L}\left[P_N^{\vec{\mu}}\right]=\mathcal{L}_\tau[\tau_{\mu_1}]a_N^{\dagger\, \mu_2}a_N^{\mu_3} m_z^{N\, \mu_4}+i[H_{\rm int},P_N^{\vec{\mu}}]+\tau_{\mu_1}\mathcal{D}_{\rm bath}\left[ a_N^{\dagger\, \mu_2}a_N^{\mu_3}m_z^{N\, \mu_4}\right]\, ,
$$
where we have introduced $\mathcal{L}_\tau[X]:=i[H_\tau,X]+\mathcal{D}_\tau[X]$. The latter map $\mathcal{L}_\tau$ gives rise to a new polynomial of terms like the one in Eq.~\eqref{poly}, after decomposing $\mathcal{L}_\tau[\tau_{\mu_1}]$ into a linear combination of Pauli operators. This part is thus under control. 

We then proceed considering the interaction Hamiltonian. We can write
\begin{equation}
\label{action-micro}
\begin{split}
i\left[H_{\rm int }, P_N^{\vec{\mu}}\right]=ig_0\sqrt{\varepsilon}\left(\tau_+a_N+\tau_-a^\dagger_N\right)P_N^{\vec{\mu}}-ig_0\sqrt{\varepsilon} P_N^{\vec{\mu}}\left(\tau_+a_N+\tau_-a^\dagger_N\right)\, .
\end{split}
\end{equation}
All four terms above can be decomposed into a linear combination of monomials of the form given in Eq.~\eqref{poly}. The operators of the central spin can simply be multiplied together and decomposed into Pauli matrices. Moreover, we have that terms like $a_NP_N^{\vec{\mu}}$ or $P_N^{\vec{\mu}}a_N^\dagger$ are close to the monomials in Eq.~\eqref{poly} even though they do not possess the correct ordering of the operators.  To reinstate the ordering fixed by our convention in Eq.~\eqref{poly}, one would have to commute the operator $a_N$ ($a_N^\dagger$) through all the $a_N^\dagger$ ($a_N$) appearing in $P_N^{\vec{\mu}}$.  Each of these commutators generates a term proportional to $m_z^N$. The latter should also be moved to its position to reconstruct polynomials of the correct form. The commutator of $m_z^N$ with $a_N$ ($a_N^\dagger$) gives again the operator $a_N$ ($a_N^\dagger$) with, however, an additional rescaling $1/N$. As such, moving $m_z^N$ can be done safely, as it generates polynomials which are of order $O(1/N)$.  With these considerations, we can conclude that all the terms appearing in Eq.~\eqref{action-micro} can be rewritten as a linear combination of the monomials in Eq.~\eqref{poly}. 
Moreover, considering the Hamiltonian $\tilde{H}_{\rm int}$ on the spin-boson model, we see that 
\begin{equation}
\label{action-emerg}
i\left[\tilde{H}_{\rm int }, P^{\vec{\mu}}\right]=ig_0\sqrt{\varepsilon}\left(\tau_+a+\tau_-a^\dagger\right)P^{\vec{\mu}}-ig_0\sqrt{\varepsilon} P^{\vec{\mu}}\left(\tau_+a+\tau_-a^\dagger\right)\, .
\end{equation}
Due to Lemma \ref{QCLT}, we thus have that when considering limits, as in Eq.~\eqref{limit-generator}, for the operator $i[H_{\rm int}, P_N^{\vec{\mu}}]$ in Eq.~\eqref{action-micro}, we obtain the operator $i[\tilde{H}_{\rm int}, P^{\vec{\mu}}]$.

We are thus left with the part of the generator associated with $\mathcal{D}_{\rm bath}$. This only acts nontrivially on bath-spin operators, so we solely focus  on the latter. We shall make use of the following relation
$$
\mathcal{D}_{\rm bath}[AB]=\mathcal{D}_{\rm bath}[A]B+A\mathcal{D}_{\rm bath}[B]+\mathcal{K}[A,B]\, , 
$$
where 
$$
\mathcal{K}[A,B]= \sum_{\alpha=\pm}\sum_{k=1}^N \left[j_\alpha^{(k)\, {\dagger}},A\right]\left[B,j_\alpha^{(k)}\right]\, ,\qquad \mbox{ with } j_-=\sqrt{\Gamma_\downarrow}\sigma_-\, , \quad \mbox{ and } \quad j_+=\sqrt{\Gamma_\uparrow}\sigma_+\, .
$$
With this expression, we observe that 
\begin{equation}
\label{termsDbath}
\begin{split}
\mathcal{D}_{\rm bath}\left[a_N^{\dagger\, \mu_2}a_N^{\mu_3} m_z^{N\, \mu_4}\right]=\mathcal{D}_{\rm bath}\left[a_N^{\dagger\, \mu_2} a_N^{\mu_3}\right]m_z^{N\, \mu_4}+a_N^{\dagger\, \mu_2}a_N^{\mu_3}\mathcal{D}_{\rm bath}\left[m_z^{N\, \mu_4}\right]
+\mathcal{K}\left[ a_N^{\dagger\, \mu_2}a_N^{\mu_3}, m_z^{N\, \mu_4}\right]\, .
\end{split}
\end{equation}
We now show that the last term provides monomials of the form in Eq.~\eqref{poly} but suppressed by, at least, a factor $1/N$. This can be seen as follows. Using the linearity of the commutator we can write the last term above as 
\begin{equation}
\label{first-K-term}
\begin{split}
\mathcal{K}\left[ a_N^{\dagger\, \mu_2}a_N^{\mu_3},m_z^{N\, \mu_4}\right]
&=\sum_{\alpha=\pm}\sum_{k=1}^N  a_N^{\dagger\, \mu_2}\left[j_\alpha^{(k)\, {\dagger}},a_N^{\mu_3}\right]\left[m_z^{N\, \mu_4},j_\alpha^{(k)}\right]\\
&+\sum_{\alpha=\pm}\sum_{k=1}^N \left[j_\alpha^{(k)\, {\dagger}}, a_N^{\dagger\, \mu_2}\right] a_N^{\mu_3}\left[m_z^{N\, \mu_4},j_\alpha^{(k)}\right]\, .
\end{split}
\end{equation}
Now we focus on the first term on the right-hand-side of the above equation. We observe that only $\sigma_+^{(k)}$ does not trivially commute with $a_N$ and we further expand the commutators to find
\begin{equation}
\begin{split}
\sum_{k=1}^N a_N^{\dagger\, \mu_2}\left[j_-^{(k)\, {\dagger}},a_N^{\mu_3}\right]\left[m_z^{N\, \mu_4},j_-^{(k)}\right]=\sum_{k=1}^N\sum_{\ell_1=0}^{\mu_3}\sum_{\ell_2=0}^{\mu_4}a_N^{\dagger\, \mu_2}a_N^{\ell_1}\left[j_-^{(k)\, {\dagger}},a_N\right]a_N^{\mu_3-1-\ell_1}m_z^{N\, \ell_2}\left[m_z^{N},j_-^{(k)}\right]m_z^{N\, \mu_4-1-\ell_2}=\\
=-\frac{2\Gamma_\downarrow}{N\sqrt{N \varepsilon}} \sum_{k=1}^N\sum_{\ell_1=0}^{\mu_3}\sum_{\ell_2=0}^{\mu_4}a_N^{\dagger\, \mu_2}a_N^{\ell_1}\sigma_z^{(k)}a_N^{\mu_3-1-\ell_1}m_z^{N\, \ell_2}\sigma_-^{(k)}m_z^{N\, \mu_4-1-\ell_2}\, .
\end{split}
\end{equation}
This is already enough to see that this term vanishes under expectation. Indeed, we can commute $\sigma_-^{(k)}$ through and bring it in front of $\sigma_z^{(k)}$. The product of the two is proportional to $\sigma_-^{(k)}$ and, using the sum over $k$ and the factor $1/\sqrt{N\varepsilon}$ we can form again an operator $a_N$. However, the resulting monomial would still be suppressed by the factor $1/N$ appearing in front of the summations. Commuting $\sigma_-^{(k)}$ through the terms in between this operator  and $\sigma_z^{(k)}$ generates additional terms. In particular the commutator $[m_z^{N},\sigma_-^{(k)}]\propto \sigma_-^{(k)}/N$. So that, we still have to do the operation of bringing this $\sigma_-^{(k)}$ in front of $\sigma_z^{(k)}$ but these terms are even more suppressed than the one previously discussed. An analogous argument applies also to the second term in Eq.~\eqref{first-K-term}. 
We can thus conclude that 
$$
\mathcal{K}\left[\left(a_N^\dagger\right)^{\mu_2}(a_N)^{\mu_3},\left(m_z^{N}\right)^{\mu_4}\right]=O\left(\frac{1}{N}\right)\, .
$$
We now go back to equation Eq.~\eqref{termsDbath} and consider the second term appearing on the right-hand-side. We want to show that this is also a $O(1/N)$ term. Since the operators $a_N,a_N^\dagger$ in the product are untouched we can just consider the term involving the average operators $m_z^N$.  We observe that 
\begin{equation}
\mathcal{D}_{\rm bath}\left[m_z^{N\, \mu_4}\right]=\mathcal{D}_{\rm bath}\left[m_z^{N}\right]m_z^{N\, \mu_4-1}+m_z^{N}\mathcal{D}_{\rm bath}\left[m_z^{N\, \mu_4-1}\right]+\mathcal{K}\left[m_z^{N},m_z^{N\, \mu_4-1}\right]\, ,
\end{equation}
and we make the last term explicit as 
$$
\mathcal{K}\left[m_z^{N},m_z^{N\, \mu_4-1}\right]=\frac{1}{N^2}\sum_{\alpha=\pm}\sum_{k=1}^N \sum_{\ell=0}^{\mu_4-2}\left[j_\alpha^{(k)\, {\dagger}},\sigma_z^{(k)}\right]m_z^{N\, \ell }\left[\sigma_z^{(k)},j_\alpha^{(k)}\right]m_z^{N\, \mu_4-2-\ell}\, .
$$
The above relation shows that the term $\mathcal{K}\left[m_z^{N},m_z^{N\, \mu_4-1}\right]$ is a sum of products of average operators which are however suppressed by an extra factor $1/N$. These are thus terms which are vanishing, in the thermodynamic limit, under any possible expectation over the state $\Omega_{\rm SS}$ of the type in Eq.~\eqref{limit-generator}. This is general and holds for any power in the second entry. As such, we can iterate the argument to show that 
$$
\mathcal{D}_{\rm bath}\left[m_z^{N\, \mu_4}\right]=\sum_{\ell=0}^{\mu_4-1}m_z^{N\, \ell}\mathcal{D}_{\rm bath}\left[m_z^N\right]m_z^{N\, \mu_4-1-\ell}+O\left(\frac{1}{N}\right)\, .
$$
The first term on the right hand side of the above equation is a product of average operators and, thus, in the large $N$ limit these converge to their expectation value. Since $\Omega_{\rm SS}\left(\mathcal{D}_{\rm bath}\left[m_z^N\right]\right)=0$ given that the state is stationary for such operators, we have that this term converges to zero with an order $1/N$, since it is also a product state. Moreover, this convergence cannot be changed by further actions of the total generator on this term, since the commutator with $H_{\rm int}$ would provide a polynomial term suppressed by a factor $1/N$ while a further application of  $\mathcal{D}_{\rm bath}$ would not change the previous argument. We thus have $\mathcal{D}_{\rm bath}\left[m_z^{N\, \mu_4}\right]=O(1/N)$ under expectation. 

We then finally consider the first term on the right-hand-side of Eq.~\eqref{termsDbath}. We have that 
\begin{equation}
\mathcal{D}_{\rm bath}\left[a_N^{\dagger\, \mu_2}a_N^{\mu_3}\right]=\mathcal{D}_{\rm bath}\left[a_N^{\dagger\, \mu_2}\right]a_N^{\mu_3}+a_N^{\dagger\, \mu_2}\mathcal{D}_{\rm bath}\left[a_N^{\mu_3}\right]+\mathcal{K}\left[a_N^{\dagger\, \mu_2},a_N^{\mu_3}\right]\, .
\end{equation}
Now, for the first and the second terms, for which the correction $\mathcal{K}$ is identically zero we have that 
$$
\mathcal{D}_{\rm bath}\left[a_N^{\dagger\, \mu_2}\right]=-\mu_2\frac{\Gamma_+}{2}a_N^{\dagger\, \mu_2}\, ,\quad \mathcal{D}_{\rm bath}\left[a_N^{ \mu_3}\right]=-\mu_3\frac{\Gamma_+}{2}a_N^{\mu_3}\, . 
$$
For the term in $\mathcal{K}$ we have
$$
\mathcal{K}\left[a_N^{\dagger\, \mu_2},a_N^{\mu_3}\right]=\sum_{\alpha=\pm}\sum_{k=1}^N \sum_{\ell_1=0}^{\mu_2-1}\sum_{\ell_2=0}^{\mu_3-1}a^{\dagger\, \ell_1}\left[j_\alpha^{(k)\, {\dagger}},a_N^{\dagger}\right]a_N^{\dagger\, \mu_2-1-\ell_1}a_N^{\ell_2}\left[a_N,j_\alpha^{(k)}\right]a_N^{\mu_3-1-\ell_2}\, .
$$
We can bring together the two commutator and outside the summations over $\ell_1,\ell_2$. This operation generates additional corrections which are however suppressed, at least as $1/N$ for $N\to\infty$. We can thus compute 
$$
\mathcal{K}\left[a_N^{\dagger\, \mu_2},a_N^{\mu_3}\right]=\frac{\Gamma_\uparrow}{\varepsilon}\mu_2\mu_3 a_N^{\dagger\, \mu_2-1}   a_N^{\mu_3-1}+O(1/N)\, .
$$
Using similar manipulations, we can check that the bosonic generator $\tilde{\mathcal{D}}$ when acting on a product $a^{\dagger \, \mu_2}a^{\mu_3}$ gives
$$
\mathcal{D}\left[a^{\dagger \, \mu_2}a^{\mu_3}\right]=-(\mu_2+\mu_3)\frac{\Gamma_+}{2}a^{\dagger \, \mu_2}a^{\mu_3}+\frac{\Gamma_\uparrow}{\varepsilon}\mu_2\mu_3 a^{\dagger\, \mu_2-1}   a^{\mu_3-1}\, . 
$$
This shows that considering the limit in Eq.~\eqref{limit-generator}, the operator $\mathcal{D}_{\rm bath}[P_N^{\vec{\mu}}]$ converges to the operator $\tilde{\mathcal{D}}[P^{\vec{\mu}}]$ and thus concludes the proof. \qed 
\\

\begin{corollary}
\label{Cor1}
Assume the following
\begin{equation*}
\begin{split}
&{\rm (A1)}\qquad \lim_{N\to\infty}\varphi\otimes \Omega_{\rm SS}\left(P_N^{\vec{\nu}}e^{t\mathcal{L}}[P_N^{\vec{\mu}}]P_N^{\vec{\eta}}\right)=\sum_{k=0}^\infty\frac{t^k}{k!}\lim_{N\to\infty}\varphi\otimes \Omega_{\rm SS}\left(P_N^{\vec{\nu}}{\mathcal{L}^k}[P_N^{\vec{\mu}}]P_N^{\vec{\eta}}\right),\\
&{\rm (A2)}\qquad \varphi\otimes\tilde{\Omega}_\beta\left(P^{\vec{\nu}}e^{t\tilde{\mathcal{L}}}\left[P^{\vec{\mu}}\right]P^{\vec{\eta}}\right)=\sum_{k=0}^\infty \frac{t^k}{k!}\varphi\otimes \tilde{\Omega}_{\beta}\left(P^{\vec{\nu}}\tilde{\mathcal{L}}^{k}\left[P^{\vec{\mu}}\right]P^{\vec{\eta}}\right) \, \mbox{ converges uniformly in $t$, for $t$ in compacts},\\
\end{split}
\end{equation*}
for any monomial $P_N^{\vec{\nu}},P_N^{\vec{\mu}},P_N^{\vec{\eta}}$ and $P^{\vec{\nu}},P^{\vec{\mu}},P^{\vec{\eta}}$. We then have that 
$$
\lim_{N\to\infty}\varphi\otimes\Omega_{\rm SS}\left(P_N^{\vec{\nu}}e^{t\mathcal{L}}[P_N^{\vec{\mu}}]P_N^{\vec{\eta}}\right)=\varphi\otimes \tilde{\Omega}_{\beta}\left(P^{\vec{\nu}}e^{t\tilde{\mathcal{L}}}[P^{\vec{\mu}}]P^{\vec{\eta}}\right)\, .
$$
\end{corollary}

{\it Proof:} As shown in the proof of Theorem \ref{Theo1}, $\mathcal{L}[P_N^{\vec{\mu}}]$ is a linear combination of  monomials as in Eq.~\eqref{poly}, plus corrections of order $O(1/N)$. The further action of the generator $\mathcal{L}$ on terms of order $O(1/N)$ cannot restore convergence of these terms to finite polynomials. Combining these two observations, we have that 
$$
\mathcal{L}^{k}[P_N^{\vec{\mu}}]\quad \longrightarrow\quad  \tilde{\mathcal{L}}^k\left[P^{\vec{\mu}}\right]\, ,
$$
where the arrow denotes convergence under any possible expectation value as in Eq.~\eqref{limit-generator}. 

We can now exploit our assumptions. Due to assumption (A1), we can write that 
$$
\lim_{N\to\infty}\varphi\otimes \Omega_{\rm SS}\left(P_N^{\vec{\nu}}e^{t\mathcal{L}}\left[P_N^{\vec{\mu}}\right]P_N^{\vec{\eta}}\right)=\sum_{k=0}^\infty \frac{t^k}{k!}\lim_{N\to\infty}\varphi\otimes \Omega_{\rm SS}\left(P_N^{\vec{\nu}}\mathcal{L}^{k}\left[P_N^{\vec{\mu}}\right]P_N^{\vec{\eta}}\right)\, ,
$$
and exploiting our observation above, we have 
$$
\lim_{N\to\infty}\varphi\otimes \Omega_{\rm SS}\left(P_N^{\vec{\nu}}e^{t\mathcal{L}}\left[P_N^{\vec{\mu}}\right]P_N^{\vec{\eta}}\right)=\sum_{k=0}^\infty \frac{t^k}{k!}\varphi\otimes \tilde{\Omega}_{\beta}\left(P^{\vec{\nu}}\tilde{\mathcal{L}}^{k}\left[P^{\vec{\mu}}\right]P^{\vec{\eta}}\right)=\varphi\otimes\tilde{\Omega}_\beta\left(P^{\vec{\nu}}e^{t\tilde{\mathcal{L}}}\left[P^{\vec{\mu}}\right]P^{\vec{\eta}}\right)\, ,
$$
where we also exploited assumption (A2), essentially  guaranteeing that the series expansion exists. 
\qed

\section*{IV. Additional interaction terms}
In this Section, we briefly show that with the approach developed in this work we can further consider additional interaction terms. 
First, we consider a Hamiltonian of the type
$$
H_{\rm 1}=\frac{\lambda_1}{\sqrt{N}}x_{\alpha_1}(S_-+S_+)+\frac{\lambda_2}{\sqrt{N}}x_{\alpha_2}(iS_+-iS_-)\, ,
$$
where $x_{\alpha_1/\alpha_2}$ are generic central spin operators. 
This term can be treated exactly in the same way as done for the Hamiltonian in the main text [cf.~Eq.~\eqref{H_int_TL}] and becomes, in the thermodynamic limit, 
$$
H_1\to \tilde{H}_1=\sqrt{\varepsilon}\left[\lambda_1x_{\alpha_1}(a+a^\dagger)+\lambda_2x_{\alpha_2}(ia^\dagger -ia)\right]\, .
$$
Moreover, it is possible to account for interactions among bath spins. For instance, essentially for the same reasoning above, we have that the Hamiltonian 
$$
H_2=\lambda_3\frac{S_+S_+}{N}+\lambda_4\frac{S_-S_-}{N}+\lambda_5\frac{S_+S_-}{N}\, ,
$$
in the thermodynamic limit becomes (under any possible expectation with our reference state $\Omega_{\rm SS}$)
$$
H_2\to \tilde{H}_2=\varepsilon \left[\lambda_3 (a^\dagger)^2+\lambda_4 a^2 +\lambda_5 a^\dagger a\right]\, .
$$
In the same way, it is also possible to consider higher-order polynomials in the terms $S_\pm$, with appropriate rescalings. As a different example, we consider the term 
$$
H_3=\lambda_6 \frac{S_zS_z}{N}\, ,
$$
where we have defined $S_z=\sum_{k=1}^N\sigma_z^{(k)}$. In order to understand, to which emergent Hamiltonian the above term gives rise, we consider the commutator
$[H_3,a_N^\dagger]$. 
We have that 
$$
[H_3,a_N^\dagger]=2\lambda_6\left(\frac{S_z}{N}a_N^\dagger +a_N^\dagger \frac{S_z}{N}\right)\, .
$$
Considering that,   $S_z/N\to -\varepsilon$ and $a_N^\dagger \to a^\dagger$, under any possible expectation taken with the  reference state, we have that 
$$
H_3\to \tilde{H}_3=-4\varepsilon \lambda_6 a^\dagger a\, .
$$
The above observations can be formalized by a straightforward adaptation of the proof of Theorem 1.

\newpage

\section*{V. Mean-Field Regime}
In this Section, we prove Theorem \ref{Theo2} which considers the behavior of the system in the regime $g=g_0/N$. \\ 

{\it Proof of Theorem \ref{Theo2}:} We start by defining the generator $\mathcal{L}_{\rm bath}$ as the map
$\mathcal{L}_{\rm bath}[X]:=i[H_{\rm bath}, X]+\mathcal{D}_{\rm bath}[X]$, where $H_{\rm bath}=\sum_\alpha h_\alpha \sum_{k=1}^N\sigma_\alpha^{(k)}$ is a noninteracting bath-spin Hamiltonian. Following the same steps reported in the proof of Lemma \ref{local-bathspin}, using $\mathcal{L}_{\rm bath}$ instead of $\mathcal{D}_{\rm bath}$, it is straightforward to show that
$$
\left\|e^{t\mathcal{L}}[m_\alpha^N]-e^{t\mathcal{L}_{\rm bath}}[m_\alpha^N]\right\|\le \frac{4\left|g_0\right|t}{N}\, , \qquad \mbox{ and }\qquad  \left\|e^{t\mathcal{L}}[\sigma_\alpha^{(k)}\sigma_\alpha^{(h)}]-e^{t\mathcal{L}_{\rm bath}}[\sigma_\alpha^{(k)}\sigma_\alpha^{(h)}]\right\|\le \frac{8\left|g_0\right|t}{N}\, .
$$
The second bound is true also when $k=h$ since in that case $\sigma_\alpha^{(k)}\sigma_\alpha^{(h)}$ is equal to the identity and the difference would actually be exactly zero. Exploiting this, we also have that 
$$
\left\|e^{t\mathcal{L}}[m_\alpha^{N\, 2}]-e^{t\mathcal{L}_{\rm bath}}[m_\alpha^{N\, 2}]\right\|\le \frac{8\left|g_0\right|t}{N}\, .
$$
Now, we define the expectation 
$
m_\alpha(t):=\lim_{N\to\infty}\varphi\otimes\Omega\left(e^{t\mathcal{L}}[m_\alpha^N]\right)\, ,
$
which, due to the above result is given by  
$
m_\alpha(t)=\lim_{N\to\infty}\varphi\otimes \Omega\left(e^{t\mathcal{L}_{\rm bath}}[m_\alpha^N]\right)
$.
Moreover, we can also show that 
$$
\lim_{N\to\infty}\varphi\otimes\Omega\left(e^{t\mathcal{L}}[m_\alpha^{N\, 2}]\right)=\lim_{N\to\infty}\varphi\otimes\Omega\left(e^{t\mathcal{L}_{\rm bath} }[m_\alpha^{N\, 2}]\right)\, .
$$
Furthermore, we have that 
$$
e^{t\mathcal{L}_{\rm bath}}[m_\alpha^{N\, 2}]=\frac{1}{N^2}\sum_{k,h=1}^Ne^{t\mathcal{L}_{\rm bath}}[\sigma_\alpha^{(k)}\sigma_\alpha^{(h)}]=\frac{1}{N}+\frac{1}{N^2}\sum_{k\neq h}e^{t\mathcal{L}_{\rm bath}}\left[\sigma_\alpha^{(k)}\right]e^{t\mathcal{L}_{\rm bath}}\left[\sigma_\alpha^{(h)}\right]\, ,
$$
where in the second equality we exploited that the generator $\mathcal{L}_{\rm bath}$ acts independently on the different bath spins. Taking the expectation, we thus have that 
$
\lim_{N\to\infty}\varphi\otimes\Omega\left(e^{t\mathcal{L}}[m_\alpha^{N\, 2}]\right)=m_\alpha^2(t)
$,
and putting all these considerations together we have that $\lim_{N\to\infty}\mathcal{E}_\alpha(t)=0=\lim_{N\to\infty}\mathcal{E}^{\rm bath}_\alpha(t)$, where
$$
\mathcal{E}_\alpha(t):=\varphi\otimes \Omega\left(e^{t\mathcal{L}}\left[\left(m_\alpha^{N}-m_\alpha(t)\right)^2\right]\right)\, , \qquad \mathcal{E}_\alpha^{\rm bath}(t):=\varphi\otimes \Omega\left(e^{t\mathcal{L}_{\rm bath}}\left[\left(m_\alpha^{N}-m_\alpha(t)\right)^2\right]\right)\, .
$$
Since the generator $\mathcal{L}_{\rm bath}$ acts independently on the different bath spins, it is then straightforward to show that the scalars $m_\alpha(t)$ obey (linear) mean-field equations. 

We now turn our attention to operators of the central spin. We define the (time-dependent) generator $\mathcal{L}_\tau(t)$ as 
$$
\mathcal{L}_\tau(t)[X]=i\left[H_\tau+g_0\left(m_-(t)\tau_++m_+(t)\tau_-\right), X\right]+\mathcal{D}_\tau[X]\, ,
$$
acting nontrivially only on central spin operators. With this, we introduce the propagator $\Lambda_{t,s}$ as
$$
\frac{\rm d}{{\rm d}t}\Lambda_{t,s}[X]=\Lambda_{t,s}\circ \mathcal{L}_\tau(t)[X]\, ,\qquad \mbox{ and } \qquad \frac{\rm d}{{\rm d}s}\Lambda_{t,s}[X]=-\mathcal{L}_\tau(s)\circ \Lambda_{t,s}[X]\, ,
$$
and we show that, in fact, it implements the dynamics of central spin operators, in the so-called weak-operator topology. This is equivalent to showing that 
$$
\lim_{N\to\infty }\varphi\otimes\Omega\left(A^\dagger e^{t\mathcal{L}}[C]B\right)=\varphi\otimes\Omega\left(A^\dagger \Lambda_{t,0}[C]B\right)=\varphi\left(A^\dagger \Lambda_{t,0}[C]B\right)\, ,
$$
$\forall A,B,C$ operators of the central spin. The second equality comes from the fact that inside the expectation  in the middle all operators act nontrivially only on the central spin. First, we write, similarly to what done for Lemma \ref{local-bathspin}, 
$$
e^{t\mathcal{L}}[C]-\Lambda_{t,0}[C]=\sum_{\alpha=x,y}\int_0^t {\rm d}s \, e^{s\mathcal{L}}\left[i\frac{g_0}{2}\left(m_\alpha^N-m_\alpha(s)\right)\left[\tau_\alpha,\Lambda_{t,s}[C]\right]\right]\, ,
$$
where we have rewritten the terms $m_\pm(t)$ and $S_\pm/N$ in terms of the real expectations $m_\alpha(t)$ and the Hermitean average operators $m_\alpha^N$. Next, we take the generic expectation above and use Cauchy-Schwarz inequality to write 
$$
I(t):=\left|\varphi\otimes \Omega\left(A^\dagger \left[e^{t\mathcal{L}}[C]-\Lambda_{t,0}[C]\right]B\right)\right|\le \|B\|\|C\||g_0| \sum_{\alpha=x,y} \int_0^t {\rm d} s \sqrt{\varphi\otimes \Omega\left(A^\dagger e^{s\mathcal{L}}\left[\left( m_\alpha^N-m_\alpha(s)\right)^2\right]A\right)}\, .
$$
Taking the limit, we find 
$
\lim_{N\to\infty}I(t) =0
$, due to the fact that we can substitute $\mathcal{L}$ with $\mathcal{L}_{\rm bath}$, that $A$ only acts on the central spin and that $\lim_{N\to\infty}\mathcal{E}^{\rm bath}_\alpha(t)=0$.  \qed

\end{document}